\documentclass[12pt]{article}
\usepackage{eurosym}
\usepackage{amsmath, amssymb}
\usepackage{amsfonts, lscape, rotating}
\usepackage{amsmath}
\usepackage{amsmath, array}
\usepackage{amsfonts, titlesec}
\usepackage{amssymb}
\usepackage{graphicx}
\usepackage{chngcntr}
\usepackage{pstricks, pstricks-add, pgfplots, tikz}
\usepackage{pst-node, verbatim, textcomp}
\usepackage[miktex]{gnuplottex}
\usepackage{tikz}
\usepackage{amsmath, amssymb}
\usepackage{etex}
\usepackage{pstricks}
\usepackage{pst-node, verbatim}
\usepackage{amsfonts, lscape}
\usepackage{amsmath}
\usepackage{amsmath}
\usepackage{amsfonts}
\usepackage{amssymb}
\usepackage{graphicx}
\usepackage{pstricks, pstricks-add, pgfplots, tikz}
\usepackage{pst-node, verbatim, textcomp}
\usepackage[miktex]{gnuplottex}
\usepackage{tikz}
\usepackage{pgfplots}
\usepackage{setspace}
\usepackage{xr}

\newtheorem{theorem}{Theorem}

\newtheorem{definition}[theorem]{Definition}

\newtheorem{proposition}[theorem]{Proposition}

\renewcommand{\Pr}{\mathrm{P}}
\usetikzlibrary{plotmarks}
\usetikzlibrary{patterns}
\usepgfplotslibrary{external}
\pgfplotsset{compat=newest} 
\usepgfplotslibrary{fillbetween}
\newrgbcolor{lightblue}{0.9 0.9 1}
\newrgbcolor{lightred}{1 0.9 0.9}
\newrgbcolor{lightgreen}{0.9 1 0.9}
\usepgfplotslibrary{groupplots}
\usetikzlibrary{pgfplots.groupplots}
\usetikzlibrary{plotmarks}
\usetikzlibrary{patterns}
\usepgfplotslibrary{external}
\pgfplotsset{compat=newest} 
\newenvironment{proof}[1][Proof]{\noindent\textbf{#1.} }{\ \rule{0.5em}{0.5em}}
\makeatletter
\def\nobreakhline{  \noalign{\ifnum0=`}\fi
\penalty\@M
\futurelet\@let@token\LT@@nobreakhline}
\def\LT@@nobreakhline{  \ifx\@let@token\hline
\global\let\@gtempa\@gobble
\gdef\LT@sep{\penalty\@M\vskip\doublerulesep}  \else
\global\let\@gtempa\@empty
\gdef\LT@sep{\penalty\@M\vskip-\arrayrulewidth}  \fi
\ifnum0=`{\fi}  \multispan\LT@cols
\unskip\leaders\hrule\@height\arrayrulewidth\hfill\cr
\noalign{\LT@sep}  \multispan\LT@cols
\unskip\leaders\hrule\@height\arrayrulewidth\hfill\cr
\noalign{\penalty\@M}  \@gtempa}
\makeatother

\begin{document}

\title{Hours worked and the U.S.\ distribution of real annual earnings
1976--2019}
\author{Ivan Fern\'andez-Val\thanks{%
\ Boston University.} \and Aico van Vuuren\thanks{%
\ University of Groningen and Gothenburg University} \and Francis Vella%
\thanks{%
\ Georgetown University} \and Franco Peracchi\thanks{%
University of Rome, Tor Vergata. \newline
We are very grateful to the editor Thierry Magnac and the referees for their
very helpful comments.}}
\maketitle

\begin{abstract}
\noindent We examine the impact of annual hours worked on annual earnings by
decomposing changes in the real annual earnings distribution into
composition, structural and hours effects. We do so via a nonseparable
simultaneous model of hours, wages and earnings. Using the Current
Population Survey for the survey years 1976--2019, we find that changes in
the female distribution of annual hours of work are important in explaining
movements in inequality in female annual earnings. This captures the
substantial changes in their employment behavior over this period. Movements
in the male hours distribution only affect the lower part of their earnings
distribution and reflect the sensitivity of these workers' annual hours of
work to cyclical factors.
\end{abstract}

\thispagestyle{empty} 
\addtocounter{page}{-1} \newpage

\section{Introduction}

This paper examines the role of annual hours of work in the changing
distribution of annual labor earnings in the U.S.\ from 1976 to 2019.\label%
{page:contribution1} This period experienced substantial fluctuations in the
level of annual hours of work due to an upward trend in female labor force
attachment and three substantial recessions. Identifying the impact of
annual hours worked on annual labor earnings requires accounting for the
simultaneity of hours and wages and the selection bias contaminating
observed wages.\label{page:challenge} We do so by extending the Fern\'{a}%
ndez-Val et al. (2021a), hereafter FVV, estimator for nonseparable models
with censored selection rules. Whereas FVV employ the estimated hours
equation to account for the selection bias in wages we also use it to
predict counterfactual labor earnings.

Empirical investigations of inequality typically focus on wages. Notable
examples include Katz and Murphy (1992), Murphy and Welch (1992), Juhn, et
al. (1993), Welch (2000), Autor, et al. (2008), Acemoglu and Autor (2011),
Autor, et al. (2016), and Murphy and Topel (2016). The distribution of
annual labor earnings provides a complementary description of inequality as
it incorporates the number of annual hours worked by those at different
points of the wage distribution. Changes in earnings incorporate variation
on earnings from movements in both labor supply and demand not captured in
wage inequality.\label{insert:economics} For example, did the drastic
changes in female labor supply exacerbate or mitigate the increases in wage
inequality? Have the increased education levels accompanying the rising
skill premia exacerbated earnings inequality by increasing the level of
annual hours of the more educated? Education effects may also reflect
changes in the demand for highly skilled labor. Another demand consideration
is the impact of recessions. These episodes witness decreases in labor
demand. Understanding their impact on earnings at different points of its
distribution assists policy makers in forming policies to, for example,
offset earnings losses or to encourage additional educational investment.
These issues motivate our focus on understanding the impact of hours on the
earnings distribution. We do not provide a detailed examination of the
factors generating the changes in the hours distribution.

To evaluate the sources of changes in earnings inequality we construct
counterfactual earnings distributions which exploit that earnings are the
product of annual hours and hourly wages. We decompose differences across
counterfactual distributions into four components. The first two are the
structural and composition effects identified in wage decomposition
exercises, although their interpretation here is more complicated. The
remaining two are hours effects. The extensive effect captures the impact of
movement from nonemployment to positive annual hours worked. The intensive
effect reflects the impact of changes in the annual hours of those working.
The combined effect of hours can be estimated without imposing a
\textquotedblleft rank invariance" assumption, but we employ it to separate
the hours effects.

This paper makes the following contributions. First, while most previous
studies treat annual hours worked as exogenous when evaluating their impact
on annual earnings we follow Altonji et al. (2013) and model hours as
endogenous. Second, we extend the existing decomposition methodologies by
quantifying the hours effects on the earnings distribution and inequality.%
\label{page:contribution2} Finally, we apply our methodology to data from
the Annual Social and Economic Supplement of the Current Population Survey,
or March CPS, for the survey years 1976--2019. We rely on existing evidence
on wages, estimated for the same sample and reported in Fern\'{a}ndez-Val et
al. (2020), hereafter FPVV, when interpreting our results.

Section \ref{section:literature} summarizes the relevant literature. Section %
\ref{section:data} discusses the data and Section~\ref{sec:model} presents
the modelling framework. The counterfactual distributions and decomposition
methodology are discussed in Section \ref{section:counterfactuals}.
Preliminaries with respect to the empirical investigation are provided in
Section \ref{section:preliminaries} and Section~\ref{sec:results} reports
and discusses the empirical results. Section~\ref{sec:conclusions} concludes.

\section{Literature overview}

\label{section:literature}

Gottschalk and Danziger (2005) examine the ratio of the upper and lower
deciles of the earnings distributions of working individuals using the CPS
for 1975 to 2002 and conclude it increased for males but decreased for
females. They attribute the decrease for females to increased working hours
of those at the bottom of the female wage distribution. Blau and Kahn (2009)
decompose changes in annual earnings into hours and wages effects and model
the variance of log annual earnings as the sum of the variances of the log
hourly wage and working hours plus twice their covariance. Examining data
for the U.S. and seven other countries for the mid to late 1990s they find
that the variance in working hours plays an important role in explaining the
differences in the variance of annual earnings. However, it is less
important for the U.S.. Checchi et al. (2016) use a similar approach in
examining the mean log deviation of earnings. Biewen and Pl\"{o}tze (2019)
investigate the role of working hours in the increased earnings inequality
in Germany. They follow a similar methodology to Blau and Kahn (2009) and
conclude, via the decomposition method of Dinardo et al. (1996), that
changes in working hours increased both male and female earnings inequality.

While the majority of papers treat working hours as exogenous to earnings
Altonji et al. (2013) \label{insert:altonji} estimate a semi-structural
model for the labor market behavior of male household heads with endogenous
labor market transitions, wages and working hours. Their main objective is
to evaluate the impact of shocks in, for example, unemployment on the
variance in the working hours, wages and earnings. This approach could also
assess the impact of working hours on the earnings distribution. Our
non-structural approach is complementary to theirs as we are able to
directly change the hours distribution without changing shocks in the
underlying economic model. Moreover, our nonparametric and nonseparable
model provides a more flexible approach to producing counterfactual earnings
distributions. Card and Hyslop (2021) also consider a model with an
endogenous work decision.\label{page:card} Examining three different PSID
cohorts of females they decompose changes in the variance into extensive and
intensive margins. Their dynamic model allows a consideration of both within
and between cohort variances in earnings. They conclude that the substantial
reduction in inequality for more recent cohorts is the result of increased
employment rates.

\section{Data}

\label{section:data}

\subsection{Data description}

We employ March CPS data from 1976 to 2019 for annual earnings, weeks
worked, and usual hours of work per week in the previous calendar year. We
restrict our analysis to those aged 24--65 years in the survey year. This
produces a sample of 1,933,659 males and 2,098,894 females. Annual sample
sizes range from a minimum of 30,767 males and 33,924 females in 1976 to a
maximum of 55,039 males and 59,622 females in 2001. Annual earnings are
reported as wage and salary incomes. Hourly wages are defined as annual
earnings divided by annual hours worked. Annual hours worked last year
(hereafter annual hours) are the product of weeks worked and usual hours of
work per week. The same level of annual hours can reflect different
combinations of hours and weeks and this has implications for evaluating
changes along the extensive and intensive margins. We discuss this in
Section \ref{sec:results}.

Individuals reporting zero hours worked last year generally respond as not
being in the labor force in the week of the March survey. We refer to these
individuals as nonemployed and note that the movement between years,
especially for males, could reflect unemployment.\label%
{page:unemployed_nonemployed} As hourly wages are defined as the ratio of
reported annual earnings and annual hours they are only available for those
in the labor force. This has implications for those in the Armed Forces, the
self-employed, and unpaid family workers as their annual earnings and annual
hours tend to be poorly measured. Accordingly, we confine attention to
civilian dependent employees with hourly wages and those out of the labor
force last year. This sample contains 1,676,014 males and 1,975,013 females
(respectively 86.7\% and 94.1\% of the original sample aged 24--65). The
subsample of civilian dependent employees with positive hourly wages
contains 1,450,941 males and 1,377,718 females.

There are two additional data issues. The first is the top coding of
earnings. \label{page:topcoding} This affects mean earnings but not
quantiles below the top coded values. The second is the difference between
actual and usual working hours. Although the CPS reports actual hours worked
in the survey week we use usual hours as they are closely related to annual
earnings. The two measures have similar means and their correlation is high
at 0.7 in 1976 and 0.65 in 2019.

\subsection{Descriptive statistics}

\label{ss:descriptive}

\pgfplotstableread{results_earnings_quantiles_raw_0.10_males.txt}{\resultsa} %
\pgfplotstableread{results_earnings_quantiles_raw_0.25_males.txt}{\resultsb} %
\pgfplotstableread{results_earnings_quantiles_raw_0.50_males.txt}{\resultsc} %
\pgfplotstableread{results_earnings_quantiles_raw_0.75_males.txt}{\resultsd} %
\pgfplotstableread{results_earnings_quantiles_raw_0.90_males.txt}{\resultse} %
\pgfplotstableread{results_earnings_quantiles_raw_0.10_females.txt}{%
\resultsf} %
\pgfplotstableread{results_earnings_quantiles_raw_0.25_females.txt}{%
\resultsg} %
\pgfplotstableread{results_earnings_quantiles_raw_0.50_females.txt}{%
\resultsh} %
\pgfplotstableread{results_earnings_quantiles_raw_0.75_females.txt}{%
\resultsi} %
\pgfplotstableread{results_earnings_quantiles_raw_0.90_females.txt}{%
\resultsj} \pgfkeys{/pgf/number format/.cd,1000 sep={}}

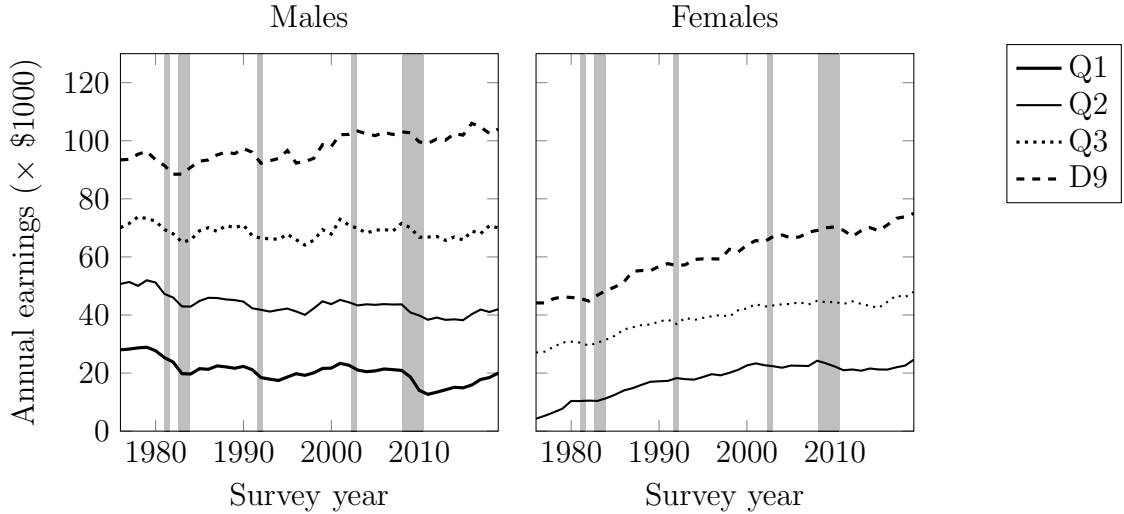
\begin{figure}[tbp]
\centering
\begin{tikzpicture}
	\begin{groupplot}
	[
	group style={%
		group size =  2 by 1,
		group name=plots,
		horizontal sep=0.5cm, 
		vertical sep=3cm, 
		xlabels at=edge bottom,
		y descriptions at=edge left,
		ylabels at=edge left,
		x descriptions at=edge bottom
	},
	set layers,cell picture=true,
	width=0.4\textwidth,
	height=0.4\textwidth,
	legend columns=1,
	xlabel = Survey year,
	ylabel = Annual earnings ($\times$ \textdollar  1000),
	ymin = 0,
	ymax = 130,
	xmin = 1976, 
	xmax = 2019,
	cycle list name=black white
	]
	\nextgroupplot[legend to name=grouplegend1, title = Males]
	\addplot[black, very thick] table[x = year, y = q1]{quantiles_males_new_1.txt};
	\addlegendentry{Q1}
	\addplot[black, thick] table[x = year, y = q2]{quantiles_males_new_1.txt};
	\addlegendentry{Q2}
	\addplot[black, very thick, dotted] table[x = year, y = q3]{quantiles_males_new_1.txt};
	\addlegendentry{Q3}
	\addplot[black, very thick, dashed] table[x = year, y = d9]{quantiles_males_new_1.txt};
	\addlegendentry{D9}
\addplot[domain=1981.0:1981.6, name path = C]{130};
\addplot[domain=1981.0:1981.6, name path = D]{0};      
\addplot[domain=1982.6:1983.92, name path = E]{130};
\addplot[domain=1982.6:1983.92, name path = F]{0};      		
\addplot[domain=1991.6:1992.25, name path = G]{130};
\addplot[domain=1991.6:1992.25, name path = H]{0};   	 
\addplot[domain=2002.25:2002.92, name path = I]{130};
\addplot[domain=2002.25:2002.92, name path = J]{0};   			
\addplot[domain=2008.09:2010.5, name path = K]{130};
\addplot[domain=2008.09:2010.5, name path = L]{0};   		
\addplot[lightgray] fill between[of=C and D];
\addplot[lightgray] fill between[of=E and F];
\addplot[lightgray] fill between[of=G and H];           
\addplot[lightgray] fill between[of=I and J];
\addplot[lightgray] fill between[of=K and L]; 		
	\nextgroupplot[title = Females]
	\addplot[black, thick]  table[x = year, y = q2]{quantiles_females_new_1.txt};
	\addplot[black, thick, dotted]  table[x = year, y = q3]{quantiles_females_new_1.txt};
	\addplot[black, very thick, dashed]  table[x = year, y = d9]{quantiles_females_new_1.txt};
\addplot[domain=1981.0:1981.6, name path = C]{130};
\addplot[domain=1981.0:1981.6, name path = D]{0};      
\addplot[domain=1982.6:1983.92, name path = E]{130};
\addplot[domain=1982.6:1983.92, name path = F]{0};      		
\addplot[domain=1991.6:1992.25, name path = G]{130};
\addplot[domain=1991.6:1992.25, name path = H]{0};   	 
\addplot[domain=2002.25:2002.92, name path = I]{130};
\addplot[domain=2002.25:2002.92, name path = J]{0};   			
\addplot[domain=2008.09:2010.5, name path = K]{130};
\addplot[domain=2008.09:2010.5, name path = L]{0};   		
\addplot[lightgray] fill between[of=C and D];
\addplot[lightgray] fill between[of=E and F];
\addplot[lightgray] fill between[of=G and H];           
\addplot[lightgray] fill between[of=I and J];
\addplot[lightgray] fill between[of=K and L]; 		
	\end{groupplot}
	\node at (plots c2r1.east) [inner sep=10pt,anchor=north, xshift= 2cm,yshift=3cm] {\ref{grouplegend1}};  
	\end{tikzpicture}
\caption{Time profile of selected quantiles of annual earnings in 2019
dollars.}
\label{fig:earnings}
\end{figure}

Figure~\ref{fig:earnings} presents the time series of various quantiles of
the annual earnings distribution for the full sample. We use the CPI to
transform these quantiles to 2019 dollars.\label{page:2016_dollars} We
present the lower quartile (Q1), the median (Q2), the upper quartile (Q3),
and the last decile (D9). Shaded vertical bars denote recessions as
identified by the NBER.

Mean female earnings increase by over 100\% and reflect the drastic changes
in both employment rates and annual hours. Female employment rates are under
75\% for almost all years of our analysis and the lowest quartile of female
earnings is nearly always zero. The quantiles we examine are generally
monotonically increasing throughout the period although there is evidence of
cyclicality. The extraordinary increase at the median of 512\% reflects the
drastic changes in the female employment behavior. There are also
substantial increases of 77\% and 70\% at Q3 and D9.

Mean male earnings decrease by 7\%. The changes are cyclical and variable.
The mean decreases substantially until around 2010 but this decrease is
subsequently partially offset. A sharp decline at the beginning of the
sample period is followed by a modest increase and a comparable decrease. A
large decrease occurs during the Great Recession. The lower quartile paints
a remarkably bleak portrait of the labor market experience at this quantile.
There are three instances of large decreases. The first is the beginning of
the 1980s, the second is the early 1990s, and the final, and most dramatic,
corresponds to the Great Recession. This produces a reduction of 53\% from
1976 to 2012. Despite the sharp rebound, earnings are 30\% lower in 2019
than in 1976. Median earnings follow a similar pattern to mean earnings.
They are variable and cyclical and decrease by 17\% over the sample period.
Earnings at Q3 and D9 show similar cyclical patterns although the reduction
for the period at Q3 is only 0.1\% while there is an increase of 11\% at D9.
The remarkably contrasting experiences of females and males is reflected in
earnings dispersion. Using the ratio (D9/Q2) as a measure of inequality,
male earnings dispersion has increased from 1.84 to 2.48, while that of
females decreased from 10.3 to 3.05. To highlight the impact of non
participation consider the corresponding quantiles of earners. Females have
gains of 90\% at Q1, 52\% at Q2, and 50\% at Q3. Males annual earnings
decrease by 15\% at Q1, 9\% at Q2, and increase by 8\% at Q3. 

Figure \ref{fig:employment} of Fern\'{a}ndez-Val et al. (2021b) shows that
the male employment rate for our data decreased from about 89.9\% in 1976 to
83.3\% in 2019. The decline is relatively steady although the larger falls
occur during cyclical downturns and the small increases follow these
recessionary periods. The female employment rate increases from 56.0\% in
1976 to 74.9\% by 2001. It falls over the remainder of our sample period
with the biggest declines in recessionary years. Beaudry et al. (2016)
conjecture that the recent decrease of employment of both males and females
reflects the diminished demand for cognitive skills which reduced the number
of high-skilled workers in well-paying jobs (see also Dillon and Veramendi,
2018).\label{page:beaudry1} This resulted in these workers taking jobs not
requiring high levels of cognitive skills and the crowding out of lower
educated workers that had held them.

%

\pgfplotstableread{results_hours_quantiles_raw_0.10_males.txt}{\resultsa} %
\pgfplotstableread{results_hours_quantiles_raw_0.25_males.txt}{\resultsb} %
\pgfplotstableread{results_hours_quantiles_raw_0.50_males.txt}{\resultsc} %
\pgfplotstableread{results_hours_quantiles_raw_0.75_males.txt}{\resultsd} %
\pgfplotstableread{results_hours_quantiles_raw_0.90_males.txt}{\resultse} %
\pgfplotstableread{results_hours_quantiles_raw_0.10_females.txt}{\resultsf} %
\pgfplotstableread{results_hours_quantiles_raw_0.25_females.txt}{\resultsg} %
\pgfplotstableread{results_hours_quantiles_raw_0.50_females.txt}{\resultsh} %
\pgfplotstableread{results_hours_quantiles_raw_0.75_females.txt}{\resultsi} %
\pgfplotstableread{results_hours_quantiles_raw_0.90_females.txt}{\resultsj} %
\pgfkeys{/pgf/number format/.cd,1000 sep={}}

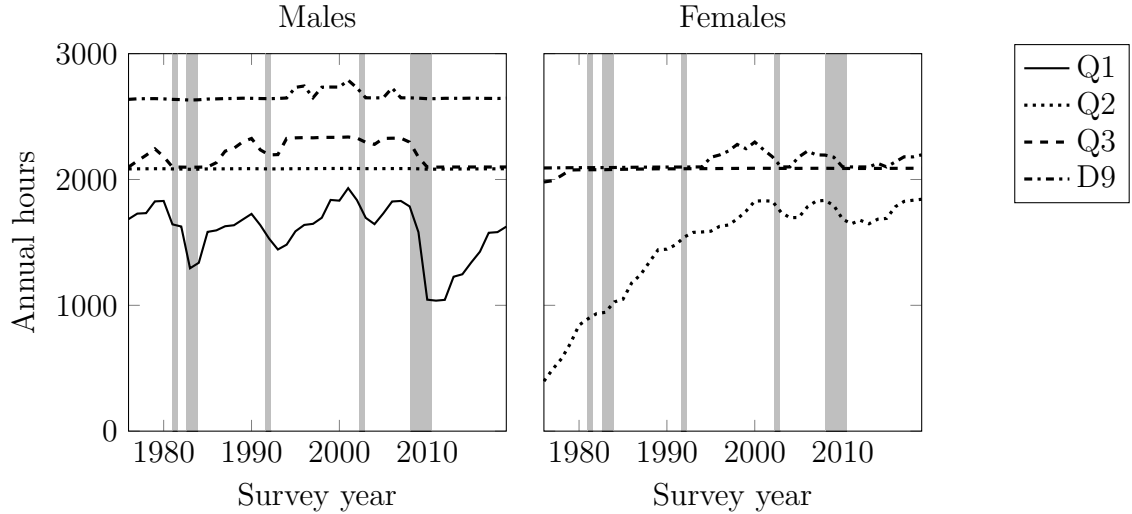
\begin{figure}[tbp]
\centering
\begin{tikzpicture}
	\begin{groupplot}
	[
	group style={%
		vertical sep = 3cm,
		group size = 2 by 1,
		group name=plots,
		horizontal sep=0.5cm, 
		xlabels at=edge bottom,
		y descriptions at=edge left,
		ylabels at=edge left,
		x descriptions at=edge bottom
	},
	set layers,cell picture=true,
	width=0.4\textwidth,
	height=0.4\textwidth,
	legend columns=1,
	xlabel = Survey year,
	ylabel = Annual hours,
	ymin = 0,
	ymax = 3000,
	xmin = 1976, 
	xmax = 2019,
	cycle list name=black white
	]	
		\nextgroupplot[legend to name=grouplegend1, title = Males]
	\addplot[black, thick] table[x = year, y = q1]{quantiles_males_hours_a.txt};
		\addlegendentry{Q1}
	\addplot[black, very thick, dotted] table[x = year, y = q2]{quantiles_males_hours_a.txt};
		\addlegendentry{Q2}
	\addplot[black, very thick, dashed] table[x = year, y = q3]{quantiles_males_hours_a.txt};
		\addlegendentry{Q3}
	\addplot[black, very thick, dashdotted] table[x = year, y = d9]{quantiles_males_hours_a.txt};
		\addlegendentry{D9}
	\addplot[domain=1981.0:1981.6, name path = C]{3000};
	\addplot[domain=1981.0:1981.6, name path = D]{0};      
	\addplot[domain=1982.6:1983.92, name path = E]{3000};
	\addplot[domain=1982.6:1983.92, name path = F]{0};      		
	\addplot[domain=1991.6:1992.25, name path = G]{3000};
	\addplot[domain=1991.6:1992.25, name path = H]{0};   	 
	\addplot[domain=2002.25:2002.92, name path = I]{3000};
	\addplot[domain=2002.25:2002.92, name path = J]{0};   			
	\addplot[domain=2008.09:2010.5, name path = K]{3000};
	\addplot[domain=2008.09:2010.5, name path = L]{0};   		
	\addplot[lightgray] fill between[of=C and D];
	\addplot[lightgray] fill between[of=E and F];
	\addplot[lightgray] fill between[of=G and H];           
	\addplot[lightgray] fill between[of=I and J];
	\addplot[lightgray] fill between[of=K and L]; 		
	\nextgroupplot[title = Females]
	\addplot[black, very thick, dotted] table[x = year, y = q2]{quantiles_females_hours_a.txt};
	\addplot[black, very thick, dashed] table[x = year, y = q3]{quantiles_females_hours_a.txt};
	\addplot[black, very thick, dashdotted] table[x = year, y = d9]{quantiles_females_hours_a.txt};
	\addplot[domain=1981.0:1981.6, name path = C]{3000};
	\addplot[domain=1981.0:1981.6, name path = D]{0};      
	\addplot[domain=1982.6:1983.92, name path = E]{3000};
	\addplot[domain=1982.6:1983.92, name path = F]{0};      		
	\addplot[domain=1991.6:1992.25, name path = G]{3000};
	\addplot[domain=1991.6:1992.25, name path = H]{0};   	 
	\addplot[domain=2002.25:2002.92, name path = I]{3000};
	\addplot[domain=2002.25:2002.92, name path = J]{0};   			
	\addplot[domain=2008.09:2010.5, name path = K]{3000};
	\addplot[domain=2008.09:2010.5, name path = L]{0};   		
	\addplot[lightgray] fill between[of=C and D];
	\addplot[lightgray] fill between[of=E and F];
	\addplot[lightgray] fill between[of=G and H];           
	\addplot[lightgray] fill between[of=I and J];
	\addplot[lightgray] fill between[of=K and L]; 	
	\end{groupplot}
	\node at (plots c2r1.east) [inner sep=10pt,anchor=north, xshift= 2cm,yshift=3cm] {\ref{grouplegend1}};  
	\end{tikzpicture}
\caption{Time profile of selected quantiles of annual hours of work using
the interval approach (interval lengths are 50 hours).}
\label{fig:hours}
\end{figure}


Figure~\ref{fig:hours} presents the distribution of annual hours for our
sample period. Male annual hours display cyclical behavior which is
particularly important at lower quantiles. There has been a large increase
in median female annual hours while higher quantiles are relatively
constant. Cyclical effects for females are weaker and are manifested as
occasional dips in otherwise upward trends. The median for males and Q3 for
females show little or no variation due to bunching in hours. Approximately
40\% of males and 30\% of females report 2080 hours in 1976. This
corresponds to 52 weeks of a 40-hour work week. Almost 50\% of males and
40\% of females report 2080 annual working hours by 2019. While this
increased bunching of annual hours has received some attention (see Bick,
Blandin and Rogerson 2021) we do not investigate it here. As the impact of
movements on the extensive margin depends on where entering workers locate
in the hours distribution, Figure \ref{fig:employment} of Fern\'andez-Val et
al. (2021b) reports that almost all working males are working at least 500
annual hours while a substantial fraction of employed females are working
less than 500 hours.


\pgfplotstableread{results_hours1_quantiles_raw_0.25_males.txt}{\resultsb} %
\pgfplotstableread{results_hours1_quantiles_raw_0.50_males.txt}{\resultsc} %
\pgfplotstableread{results_hours1_quantiles_raw_0.75_males.txt}{\resultsd} %
\pgfplotstableread{results_hours1_quantiles_raw_0.25_females.txt}{\resultsg} %
\pgfplotstableread{results_hours1_quantiles_raw_0.50_females.txt}{\resultsh} %
\pgfplotstableread{results_hours1_quantiles_raw_0.75_females.txt}{\resultsi} %
\pgfkeys{/pgf/number format/.cd,1000 sep={}}

\begin{figure}[tbp]
\centering
\begin{tikzpicture}
	\begin{groupplot}
	[
	group style={%
		group size = 2 by 1,
		group name=plots,
		horizontal sep=0.5cm, 
		xlabels at=edge bottom,
		y descriptions at=edge left,
		ylabels at=edge left,
		x descriptions at=edge bottom
	},
	set layers,cell picture=true,
	width=0.4\textwidth,
	height=0.4\textwidth,
	legend columns=1,
	xlabel = Survey year,
	ylabel = Annual hours,
	ymin = 0,
	ymax = 3000,
	xmin = 1976, 
	xmax = 2016,
	cycle list name=black white
	]	
	\nextgroupplot[legend to name=grouplegend1, title = Males]			
	\addplot[black, thick] table[x = year, y = q1]{mean_hours_quantiles_males.txt};
	\addlegendentry{Q1}
	\addplot[black, thick, dashed] table[x = year, y = q2]{mean_hours_quantiles_males.txt};
	\addlegendentry{Q2}
	\addplot[black, thick, dotted] table[x = year, y = q3]{mean_hours_quantiles_males.txt};
	\addlegendentry{Q3}
	\addplot[black, thick, dashdotted] table[x = year, y = q4]{mean_hours_quantiles_males.txt};
	\addlegendentry{Q4}	
        \addplot[domain=1981.0:1981.6, name path = C]{3000};
         \addplot[domain=1981.0:1981.6, name path = D]{0};      
         \addplot[domain=1982.6:1983.92, name path = E]{3000};
	\addplot[domain=1982.6:1983.92, name path = F]{0};      		
         \addplot[domain=1991.6:1992.25, name path = G]{3000};
         \addplot[domain=1991.6:1992.25, name path = H]{0};   	 
         \addplot[domain=2002.25:2002.92, name path = I]{3000};
         \addplot[domain=2002.25:2002.92, name path = J]{0};   			
         \addplot[domain=2008.09:2010.5, name path = K]{3000};
	\addplot[domain=2008.09:2010.5, name path = L]{0};   		
         \addplot[lightgray] fill between[of=C and D];
         \addplot[lightgray] fill between[of=E and F];
         \addplot[lightgray] fill between[of=G and H];           
         \addplot[lightgray] fill between[of=I and J];
         \addplot[lightgray] fill between[of=K and L]; 		
	\nextgroupplot[title = Females]	
	\addplot[black, thick] table[x = year, y = q1]{mean_hours_quantiles_females.txt};
	\addplot[black, thick, dashed] table[x = year, y = q2]{mean_hours_quantiles_females.txt};
	\addplot[black, thick, dotted] table[x = year, y = q3]{mean_hours_quantiles_females.txt};	
	\addplot[black, thick, dashdotted] table[x = year, y = q4]{mean_hours_quantiles_females.txt};
        \addplot[domain=1981.0:1981.6, name path = C]{3000};
         \addplot[domain=1981.0:1981.6, name path = D]{0};      
         \addplot[domain=1982.6:1983.92, name path = E]{3000};
	\addplot[domain=1982.6:1983.92, name path = F]{0};      		
         \addplot[domain=1991.6:1992.25, name path = G]{3000};
         \addplot[domain=1991.6:1992.25, name path = H]{0};   	 
         \addplot[domain=2002.25:2002.92, name path = I]{3000};
         \addplot[domain=2002.25:2002.92, name path = J]{0};   			
         \addplot[domain=2008.09:2010.5, name path = K]{3000};
	\addplot[domain=2008.09:2010.5, name path = L]{0};   		
         \addplot[lightgray] fill between[of=C and D];
         \addplot[lightgray] fill between[of=E and F];
         \addplot[lightgray] fill between[of=G and H];           
         \addplot[lightgray] fill between[of=I and J];
         \addplot[lightgray] fill between[of=K and L]; 		
	\end{groupplot}	
		\node at (plots c2r1.east) [inner sep=10pt,anchor=north, xshift= 2cm,yshift=3cm] {\ref{grouplegend1}};  
	\end{tikzpicture}
\caption{Time profile of mean annual hours of work by location in the annual
earnings distribution.}
\label{fig:hours1}
\end{figure}
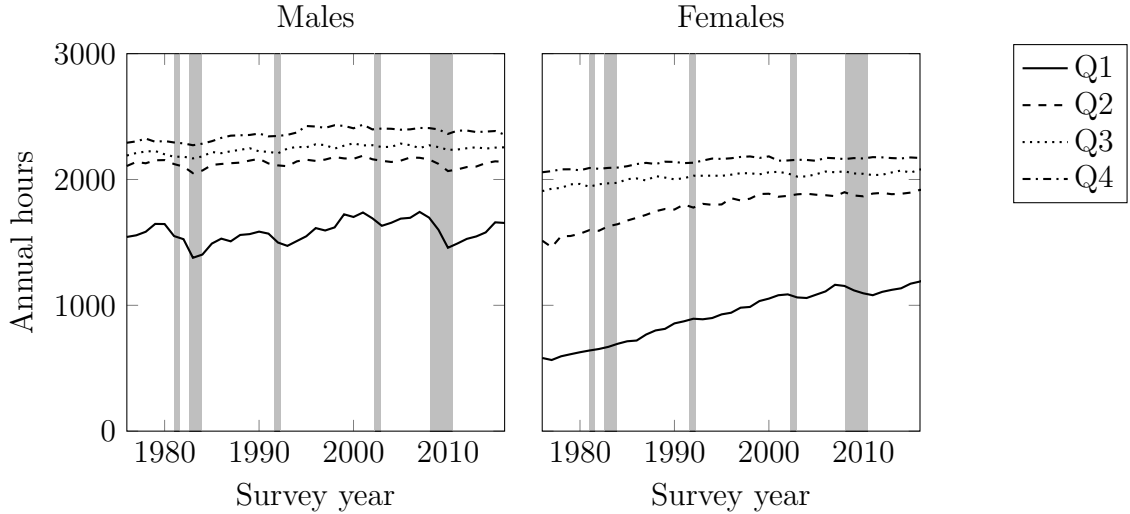

Figure~\ref{fig:hours1} presents mean hours by different quartiles of the
earnings distribution. Q1 denotes mean hours for workers with labor income
less than the lower quartile of earnings. Q2, Q3 and Q4 are defined
similarly. Male hours have not substantially increased at any point in the
earnings distribution and their cyclicality is more evident at lower
earnings. There are small increases at higher female earnings and much
larger increases at the median and below. The number of female annual hours
worked has doubled over the sample period at the lowest quartile (see also
Gottschalk and Danziger, 2005).

We also consider if movements in annual hours and employment rates reflect
notable changes in their likely determinants. Education levels, particularly
of females, increased dramatically. The large increases in those obtaining
college degrees are sometimes attributed to the decreasing selectivity of
colleges although the effects appear too small to fully explain these
observed increases (Hoxby 2009).\label{page:hoxby1} Figures \ref%
{fig:education_participation} and \ref{fig:education_hours} of Fern\'{a}%
ndez-Val et al. (2021b) show that college graduates have the highest
employment rates and average working hours over the whole sample period for
both genders while the difference between education levels is cyclical.

We also explore the relationship between annual hours and some demographic
characteristics which influence labor supply. Figure \ref%
{fig:relative_hours_married} of Fern\'{a}ndez-Val et al. (2021b) illustrates
that males work more when they are married or have children while the
opposite is observed for females. While the magnitude of these effects for
females has decreased over time, their relative ordering has remained
stable. We examined NLSY data to investigate the relationship between
working hours and cognitive skills at different periods and find a
consistently positive relationship between working hours and the AFQT test
score for both males and females.

This section highlights that the implied drastic increases in males'
earnings inequality are not accompanied by corresponding changes in annual
hours worked although the annual hours of those at lower annual earnings are
more cyclical. The drastic changes in annual earnings for females are
accompanied by substantial changes in their annual hours. Females' hours are
increasing throughout their earnings distribution with the more drastic
increases at lower earnings. The data also suggest that the nature of
relationships between an individual's observed characteristics and their
employment decision and annual hours has not changed. This supports our
modeling approach which assumes the types of individuals, as captured by
both their observed and unobserved characteristics, likely to work many
hours in each cross section do not change over the sample period.

\section{Modeling framework}

\label{sec:model}

Consider the following model describing earnings, wages and hours: 
\begin{align}
Y_{t}& =%
\begin{cases}
W_{t}H_{t} & \mbox{if}~H_{t}>0 \\ 
0 & \mbox{otherwise}%
\end{cases}%
,  \label{eq3} \\
W_{t}& =g_{t}(X_{t},E_{t}),\text{ if $H_{t}>0$},  \label{eq1} \\
H_{t}& =\max \{k_{t}(X_{t},Z_{t},V_{t}),0\},  \label{eq2}
\end{align}%
where $Y_{t}$, $W_{t}$, and $H_{t}$ denote annual earnings, hourly wages,
and annual hours of work for an individual in year $t$, respectively.
Equation (\ref{eq3}) specifies that $Y_{t}$ is the product of $W_{t}$ and $%
H_{t}$. Equation (\ref{eq1}) states that $W_{t}$ is a function of observed
individual characteristics, $X_{t}$, and unobserved characteristics, $E_{t}$%
. The function $g_{t}$ is assumed to be unknown and nonseparable with
respect to its arguments. $W_{t}$ is only observed for individuals reporting
positive annual hours. The process generating observed annual hours is
defined in \eqref{eq2}. $H_{t}$ is a function of $X_{t}$ and additional
observed characteristics, $Z_{t}$, that influence the individuals'
preference for work but not their wage. The random variable $V_{t}$
represents unobserved characteristics that affect $H_{t}$. The relationship
between these determinants and hours is represented by the unknown and
nonseparable function $k_{t}$. $H_{t}$ is censored from below at zero.


The support of random variables and vectors is denoted by calligraphic
letters; e.g., $\mathcal{XZ}_{t}$ is the joint support of $(X_{t},Z_{t})$.
Since this can depend on whether the individual belongs to the
\textquotedblleft selected population\textquotedblright\ with $H_{t}>0,$ an
asterisk indicates support in the selected population. For example, $%
\mathcal{X}_{t}^{\ast }$ is the support of $X_{t}\mid H_{t}>0$. Small
letters in parentheses indicate the support is conditional on other
variable(s) taking a particular value; e.g. $\mathcal{Z}_{t}(x)$ is the
support of $Z_{t}\mid X_{t}=x$.

Equations ~\eqref{eq1}-\eqref{eq2} comprise the model of FVV. We augment it
with (\ref{eq3}). Two assumptions are required to continue. First, the
vector $(X_{t},Z_{t})$ is independent of the vector $(E_{t},V_{t})$ in the
entire population. This excludes additional issues related to endogeneity.
This assumption may be strong for human capital variables but we do not
address this here. Second, $V_{t}$ is continuously distributed, and the
distribution of $V_{t}$ and $v\mapsto k_{t}(\cdot ,\cdot ,v)$ are strictly
increasing on the support of $V_{t}$. Given these conditions, we can
normalize $V_{t}\sim U(0,1)$.

We allow for dependence between $E_{t}$ and $V_{t}$, implying that
unobserved characteristics affecting the hourly wage are correlated with
those contributing to annual hours.\label{page:intuition_begin} This arises
in the labor supply literature (Killingsworth, 1983) where working hours are
chosen according to the maximization of a utility function, denoted by $U(c,
l)$, which depends on consumption $c$ and leisure $l$. Then $H =\arg
\max_{h}U(g_{t}(X,E)h+S,\bar H-h)$ where $S$ and $\bar H$ denote non-labor
income and total time available. The restrictions for participation are $H>0$
and $U(g_{t}(X,E)H+S,\bar H-H)>U(S,\bar H)$ implying that any (un)observed
heterogeneity in $g_{t}$ should also be in $k_{t}$.

The correlation between $E_{t}$ and $V_{t}$ produces a selection bias in
estimating $g_{t}$ over the sample for $H_{t}>0$ even when $E_{t}$ and $%
X_{t} $ are independent over the entire population. We address this via the
FVV control function approach based on the Imbens and Newey (2009) estimator
but allowing for censored selection. The appropriate variable is $V_{t}$,
which captures the \textquotedblleft rank\textquotedblright\ of the
individual's annual hours relative to observationally equivalent
individuals. $V_{t}$ is a control function since 
conditioning on $X_{t}$, $Z_{t}$ and $V_{t}$ makes selection, \emph{i.e.} $%
k_{t}(X_{t},Z_{t},V_{t})>0$, deterministic. Thus, the distribution of $E_{t}$
conditional on $X_{t}$, $Z_{t}$ and $V_{t}$ does not stochastically depend
on selection. Moreover, conditional on $V_{t}$, $E_{t}$ does not depend on $%
X_{t}$ and $Z_{t}$ in the selected population.\label{page:explain}

As $V_{t}$ is constructed using the movement in $H_{t}$ for observations $%
H_{t}>0$, it has variation conditional on $(X_{t},Z_{t}).$\label%
{insert:identification_explained} This contrasts with the propensity score,
often used as the basis of control functions in dichotomous selection
models, which is constant conditional on $(X_{t},Z_{t})$. Moreover, as
monotonicity implies $V_{t}$ equals the distribution of $H_{t}\mid
X_{t},Z_{t}$ it is identified provided $H_{t}>0$.

To distinguish between extensive and intensive margins we consider a
generalization of the model \eqref{eq3}-\eqref{eq2} where: 
\begin{align}
Y_{t}& =%
\begin{cases}
W_{t}H_{t} & \mbox{if}~k_{t}^{2}(X_{t},Z_{t},V_{t}^{2})>0, \\ 
0 & \mbox{otherwise,}%
\end{cases}
\tag{\ref{eq3}'}  \label{eq 8}
\end{align}
\begin{align}
W_{t}& =g_{t}(X_{t},E_{t})\text{ if $k_{t}^{2}(X_{t},Z_{t},V_{t}^{2})>0$}, 
\tag{\ref{eq1}'}  \label{eq 6} \\
H_{t}& =\max \{k_{t}^{1}(X_{t},Z_{t},V_{t}^{1}),0\},  \tag{\ref{eq2}'}
\label{eq 7}
\end{align}%
where $k_{t}^{1}$ and $k_{t}^{2}$ satisfy the same conditions as equation %
\eqref{eq2}, and $k_{t}^{1}(X_{t},Z_{t},V_{t}^{1})\geq
k_{t}^{2}(X_{t},Z_{t},V_{t}^{2})$. 
Here, $k_{t}^{1}$ represents the hours worked equation and $k_{t}^{2}$
captures the participation or employment equation. Model \eqref{eq3}-%
\eqref{eq2} is a special case with $k_{t}^{1}=k_{t}^{2}$ and $%
V_{t}^{1}=V_{t}^{2}$. The model \eqref{eq 8}-\eqref{eq 7} is useful for
counterfactual scenarios in which, for example, we examine how the
population from a year $t$ with a high participation rate would participate
in a year $s$ with a lower participation rate, i.e. $%
k_{t}(X_{t},Z_{t},V_{t})\geq k_{s}(X_{t},Z_{t},V_{t})$, while keeping the
hours worked equation as in year $t$. This scenario can be modeled by
setting $k_{t}^{1}=k_{t}$, $k_{t}^{2}=k_{s}$ and $V_{t}^{1}=V_{t}^{2}=V_{t}$%
. We return to this in Section \ref{section:counterfactuals}. The condition $%
V_{t}^{1}=V_{t}^{2}$ implies that for every $(X_{t},Z_{t})$ combination, $%
k_{t}^{1}(X_{t},Z_{t},V_{t})=k_{t}^{2}(X_{t},Z_{t},V_{t})+\Psi
_{t}(X_{t},Z_{t}),$ for some positive function $\Psi _{t}$. This enables an
investigation of when workers decide not to work if their \textquotedblleft
potential\textquotedblright\ working hours are lower than a certain cutoff
value $\Psi (X_{t},Z_{t})\geq 0$. It also imposes rank invariance in the
hours worked and participation equations conditional on $(X_{t},Z_{t})$
(e.g., Chernozhukov and Hansen, 2005).\label{page:rank_invariance copy(1)} %
\label{page:rank_invariance3} In the previous example, it implies the number
of hours worked does not change for those who remain working after the
counterfactual change in the participation equation.

\section{Objects of Interest}

\label{section:counterfactuals}


Our main objects of interest are the observed and counterfactual earnings
distributions. The counterfactual distributions are constructed from %
\eqref{eq3}-\eqref{eq2} and \eqref{eq 8}-\eqref{eq 7} by combining the
functions $g_{t}$ and $k_{t}$ and the distribution of $(X_{t},Z_{t},V_{t})$
from different years. 
A critical component of these distributions is the relationship between
earnings, wages and hours. For any $y>0:$ 
\begin{equation*}
\Pr (Y_{t}\leq y)=\Pr \left( W_{t}\leq \frac{y}{H_{t}}\mid H_{t}>0\right)
\Pr (H_{t}>0)+\Pr (H_{t}=0),
\end{equation*}%
where the second term reflects that the probability that $Y_{t}\leq y$
equals unity when $H_{t}=0$, and $\Pr (Y_{t}\leq 0)=\Pr (H_{t}=0).$ The same
relationship holds when we condition on $(X_{t},Z_{t},V_{t})$.

We write the observed distribution of earnings in terms of the functions of
the model. By the law of iterated probability: 
\begin{equation*}
F_{Y_{t}}(y)=\Pr [Y_{t}\leq y]=\int \Pr [Y_{t}\leq y\mid
X_{t}=x,Z_{t}=z,V_{t}=v]\,dF_{X_{t},Z_{t},V_{t}}(x,z,v).
\end{equation*}%
Using the model \eqref{eq3}-\eqref{eq2}, for any $y>0$: 
\begin{equation}
F_{Y_{t}}(y)=\int_{k_{t}(x,z,v)>0}G_{t}\left( \frac{y}{k_{t}(x,z,v)}%
,x,v\right) dF_{X_{t},Z_{t},V_{t}}(x,z,v)+\int_{k_{t}(x,z,v)\leq
0}dF_{X_{t},Z_{t},V_{t}}(x,z,v).  \label{eq:counterfactual_base}
\end{equation}%
with $G_{t}(w,x,v):=\Pr (g_{t}(x,E_{t})\leq w\mid V_{t}=v)$. We drop the
conditioning of $X_{t}$ and $Z_{t}$ in the probability of the first integral
since $E_{t}$ is independent of these variables conditional on $V_{t}$. FVV
referred to $G_{t}(w,x,v)$ as the local distribution structural function
(LDSF). It represents the distribution of potential hourly wages $%
g_{t}(x,E_{t})$ corresponding to $X_{t}=x$ conditional on $V_{t}=v$. %
%
%
%
%
%
%
%
%
%
%
FVV show that identification of the LDSF requires that the specific
combination of $x$ and $v$ being considered belongs to the support of $%
(X_{t},V_{t})$ in the selected population, \emph{i.e.} $(x,v)\in \mathcal{XV}%
_{t}^{\ast }$.\label{insert:identificiation_explanation1} This requires a
realization of $(X_{t},Z_{t},V_{t})$ equal to $(x,z,v)$ such that $%
k_{t}(x,z,v)>0$. 
The exclusion restriction $Z_{t}$ plays two roles. First, it is needed to
guarantee that $(X_{t},Z_{t})$ is independent of $(E_{t},V_{t})$. \ That is, 
$Z_{t}$ must be correctly specified to ensure the control function is
correctly estimated. Second, it expands the set $\mathcal{XV}_{t}^{\ast }$
if there is even only one value $z\in Z_{t}$ for which $k_{t}(x,z,v)$ is
positive. This expansion is useful if, for example, we are interested in
identifying the marginal distribution of $g_{t}(x,E_{t})$ in the selected or
entire populations.


%
%
%
%
%
%
%
%
%
%
%
Equation (\ref{eq:counterfactual_base}) is the basis of the counterfactual
distributions. We define a counterfactual earnings distribution as: 
\begin{equation}
\begin{split}
G_{\langle r,s,p\rangle }(y)=& \int_{k_{r}(x,z,v)>0}G_{s}\left( \frac{y}{%
k_{r}(x,z,v)},x,v\right) dF_{X_{p},Z_{p},V_{p}}(x,z,v) \\
& +\int_{k_{r}(x,z,v)\leq 0}dF_{X_{p},Z_{p},V_{p}}(x,z,v),
\end{split}
\label{eq:g_qrst}
\end{equation}%
which corresponds to the resulting earnings distribution if the hours worked
equation were as in year $r$, the distribution of explanatory variables and
control function as in year $p$, and the wage structure (LDSF) as in year $s$%
. For example, the observed distribution of earnings in year $t$\
corresponds to $G_{\langle t,t,t\rangle }=F_{Y_{t}}$.

Similarly, we can define counterfactual distributions of earnings from model %
\eqref{eq 8}-\eqref{eq 7}. In this case: 
\begin{equation}
\begin{split}
G_{\langle q,r,s,p\rangle }(y)& =\int_{k_{q}^{2}(x,z,v_{2})>0}G_{s}\left( 
\frac{y}{k_{r}^{1}(x,z,v_{1})},x,v_{1}\right)
dF_{X_{p},Z_{p},V_{p}^{1},V_{p}^{2}}(x,z,v_{1},v_{2}) \\
& +\int_{k_{q}^{2}(x,z,v_{2})\leq
0}dF_{X_{p},Z_{p},V_{p}^{1},V_{p}^{2}}(x,z,v_{1},v_{2}),
\end{split}
\label{eq:g_qrst2}
\end{equation}%
which corresponds to the resulting earnings distribution if the
participation equation were as in year $q$, hours worked equation as in year 
$r$, distribution of explanatory variables and control function as in year $%
p $, and wage structure as in year $s$. This expression yields the
counterfactual distribution \eqref{eq:g_qrst} if $k_{q}^{2}=k_{r}^{1}$ and $%
V_{r}^{1}=V_{q}^{2}$.


\subsection{Identification of Counterfactual Earnings Distributions}

\label{ss:ident}

A first step in the empirical analysis using counterfactual distributions is
to establish their nonparametric identification.\label%
{page:support_restrictions} The distributions $G_{\langle r,s,t\rangle }$
and $G_{\langle q,r,s,t\rangle }$ are identified if all the integrands are
well-defined over supports that contain the regions of integration. The
following proposition gathers support conditions that guarantee
identification and is proven in Appendix \ref{web:id} of Fern\'{a}ndez-Val
et al. (2021b).

\begin{proposition}[Identification of $G_{\langle r,s,t\rangle }$ and $%
G_{\langle q,r,s,t\rangle }$]
\label{prop:id} Under the model \eqref{eq3}--\eqref{eq2}, the distribution $%
G_{\langle r,s,p\rangle }$ in \eqref{eq:g_qrst} is identified if $\mathcal{XV%
}_{p}\cap \mathcal{XV}_{r}^{\ast }\subseteq \mathcal{XV}_{s}^{\ast }$, and $%
\mathcal{XZ}_{p}\subseteq \mathcal{XZ}_{r}$. Under the model \eqref{eq 6}-%
\eqref{eq 7}, the distribution $G_{\langle q,r,s,p\rangle }$ in %
\eqref{eq:g_qrst2} is identified if $\mathcal{XV}_{p}\cap \mathcal{XV}%
_{q}^{\ast }\subseteq \mathcal{XV}_{s}^{\ast }$, $\mathcal{XZ}_{q}^{\ast
}\cap \mathcal{XZ}_{p}\subseteq \mathcal{XZ}_{r}$, $\mathcal{XZ}%
_{p}\subseteq \mathcal{XZ}_{q}$, and $V^1_q = V^2_r$ a.s.
\end{proposition}

The conditions of Proposition \ref{prop:id} simplify when we only consider
two years for $q$, $r$, $s$ and $p$, such as $0$ and $t$, which is relevant
for the earnings decompositions in the next section. For example, we only
need $\mathcal{XV}_{t}\cap \mathcal{XV}_{0}^{\ast }\subseteq \mathcal{XV}%
_{t}^{\ast }$ and $\mathcal{XZ}_{t}\subseteq \mathcal{XZ}_{0}$ to identify $%
G_{\langle 0,t,t,t\rangle }$ and $G_{\langle 0,0,t,t\rangle }$, and $%
\mathcal{XZ}_{t}\subseteq \mathcal{XZ}_{0}$ to identify $G_{\langle
0,0,0,t\rangle }$. A sufficient condition for $\mathcal{XV}_{t}\cap \mathcal{%
XV}_{0}^{\ast }\subseteq \mathcal{XV}_{t}^{\ast }$ is that the employment
rates in year $0$, conditional on $X$, be lower than those in year $t$. For
females, this is satisfied if the year $t$ is after year $0$. A sufficient
condition for $\mathcal{XZ}_{t}\subseteq \mathcal{XZ}_{0}$ is that any $%
(x,z) $ observed in year $t$ be also observed in year $0$.

The impact of $Z$ on the plausibility of the condition $\mathcal{XV}_{p}\cap 
\mathcal{XV}_{q}^{\ast }\subseteq \mathcal{XV}_{s}^{\ast }$ is ambiguous.
Without further assumptions, $Z$ might expand each of the sets $\mathcal{XV}%
_{p}$, $\mathcal{XV}_{q}^{\ast }$ and $\mathcal{XV}_{s}^{\ast }$
differently. However, if $Z$ has sufficient enough variation to ensure $%
\mathcal{XV}=\mathcal{X}\times \lbrack 0,1]$ in each period the condition
simplifies to $\mathcal{X}_{p}\cap \mathcal{X}_{q}^{\ast }\subseteq \mathcal{%
X}_{s}^{\ast }$.

\subsection{Earnings Decompositions}

\label{ss:decompositions}


Let $Q_{\langle q,r,s,p\rangle }(\tau )$\ denote the counterfactual $\tau $%
-quantile of earnings corresponding to the left-inverse of the distribution $%
y\mapsto G_{\langle q,r,s,p\rangle }(y)$.\footnote{%
\ That is, $Q_{\langle q,r,s,p\rangle }(\tau )=\inf \{y\in \mathbb{R}\colon
G_{\langle q,r,s,p\rangle }(y)\geq \tau \}$, for $\tau \in (0,1)$.} The
definitions above allow the following decomposition of the observed change
in the $\tau $-quantile of earnings between the baseline year (labeled as
year~0) and any other year~$t$: 
\begin{align}
\frac{Q_{\langle t,t,t,t\rangle }(\tau )-Q_{\langle 0,0,0,0\rangle }(\tau )}{%
Q_{\langle 0,0,0,0\rangle }(\tau )}& =\underset{[1]}{\underbrace{\frac{%
[Q_{\langle t,t,t,t\rangle }(\tau )-Q_{\langle 0,t,t,t\rangle }(\tau )]}{%
Q_{\langle 0,0,0,0\rangle }(\tau )}}}+\underset{[2]}{\underbrace{\frac{%
[Q_{\langle 0,t,t,t\rangle }(\tau )-Q_{\langle 0,0,t,t\rangle }(\tau )]}{%
Q_{\langle 0,0,0,0\rangle }(\tau )}}}  \notag \\
& \quad +\underset{[3]}{\underbrace{\frac{[Q_{\langle 0,0,t,t\rangle }(\tau
)-Q_{\langle 0,0,0,t\rangle }(\tau )]}{Q_{\langle 0,0,0,0\rangle }(\tau )}}}+%
\underset{[4]}{\underbrace{\frac{[Q_{\langle 0,0,0,t\rangle }(\tau
)-Q_{\langle 0,0,0,0\rangle }(\tau )]}{Q_{\langle 0,0,0,0\rangle }(\tau )}}},
\label{eq:decomposition}
\end{align}%
provided that $Q_{\langle 0,0,0,0\rangle }(\tau )>0$.

Term~[1] is the \textquotedblleft extensive hours effect\textquotedblright\
capturing how individuals moving from working to non working reduce their
earnings to zero. Term~[2] is the \textquotedblleft intensive hours
effect\textquotedblright\ and measures earnings movements resulting from
shifts in the distribution of annual hours for those who work conditional on
their characteristics (i.e. $X_{t}$ and $V_{t}$). The sum of [1] and [2] is
the \textquotedblleft total hours effect\textquotedblright , $[1]+[2]=\frac{%
Q_{\langle t,t,t,t\rangle }(\tau )-Q_{\langle 0,0,t,t\rangle }(\tau )}{%
Q_{\langle 0,0,0,0\rangle }(\tau )}.$ Term ~[3] is the \textquotedblleft
structural wage effect\textquotedblright\ reflecting earnings changes from
movements in the hourly wage distribution for workers conditional on their
characteristics and captures the changing market evaluation of these
characteristics. 
Term~[4] is the \textquotedblleft composition effect\textquotedblright\ and
represents changes in the earnings distribution due to changes in the
distribution of worker characteristics.

\label{page:effects}The extensive hours effect captures the difference
between the quantiles of observed earnings in year $t$\ and those that would
prevail if the participation equation was as in year $0$. As this reflects
the mapping of the $X$ and $Z$ into participation via the hours equation for
year $0$, $k^{2}$, it corresponds to a structural effect. Due to the
identification assumptions for $G_{\langle 0,t,t,t\rangle }$, it is always
non-negative. That is, an individual with a positive probability of working
in base year $0$ should have a positive probability in year $t$, i.e. $%
k_{0}(X_{t},Z_{t},V_{t})\leq k_{t}(X_{t},Z_{t},V_{t})$. 
By the monotonicity and rank invariance assumptions, those who stop working
under the participation equation $k_{0}$ will have low levels of $V_{t}$ and
small numbers of annual hours before the change.


The intensive hours effect is the difference between the quantiles of
earnings from changing the hours equation from year~$t$\ to year $0$, while
keeping the functions $G$\ and $F_{X,Z,V}$ as in year $t$\ and the
participation equation as in year $0$. As we go from the base year to year $%
t $, this will capture differences in the hours distribution of those
working. This is also a structural effect on hours worked as it results from
changing the hours equation $k^{1}$ which maps the $X$ and $Z$ into annual
hours from $k_{t}$ to $k_{0}$. This effect is also non-negative by
construction as individuals work less under the hours worked equation of
year $0$ and we are keeping hourly wages fixed. We highlight again that
these differences may reflect either changes in weekly hours, weeks worked
or unemployment spells.

The composition effect contains the impact on earnings of changes via
movements in the composition of both the observable and unobservable
characteristics that affect wages and/or working hours. For example,
increases in female educational attainment could shift the distribution of
earnings to the right due to increases both in the hourly wages and annual
hours. Due to the nonseparable and nonparametric nature of the model, it
does not seem possible to disentangle these two effects. The higher
educational levels which increased hours, and thus higher annual earnings
(on both the extensive and intensive margin), are interpreted as composition
and not hours effects.

The structural wage effect captures how the individual's characteristics map
into the distribution of hourly wages and annual earnings. The structural
effects operating on earnings through hours are assigned either the
interpretation of extensive or intensive hours effects. Studies that
estimate parametric selection models generally attribute changes in the
values of the control function and its coefficient as selection effects.
However, FPVV show that in a nonseparable model this attribution is  not
possible. We interpret the impact of changes in the value of the control
function as a composition effect and a change in the impact of the control
function on earnings, operating through hourly wages, as a structural effect.

As with the Oaxaca-Blinder decomposition, the baseline year $0$ and the
order of the decomposition have implications for the interpretation and
magnitude of the effects. They also matter here for the plausibility of the
support conditions needed for nonparametric identification of the
counterfactual distributions involved in the decomposition. This makes it
convenient to choose the year with the lower participation as the baseline
year.

\subsection{Estimators}

The decomposition in \eqref{eq:decomposition} requires estimation of the
counterfactual distributions in equation \eqref{eq:g_qrst2} comprising
components from different years. We start by rewriting this equation as:%
\footnote{%
To rewrite the bound $k_{r}(x,z,v)>0$, we use that $F_{H_{r}\mid
X_{r},Z_{r}}(k_{r}(x,z,v)\mid x,z)>F_{H_{r}\mid X_{r},Z_{r}}(0\mid x,z)$ and 
$\Pr (k_{r}(x,z,V)\leq k_{r}(x,z,v)\mid x,z)=v$ by the assumptions on the
hours worked equation.\label{footnote:explain}} 
\begin{equation}
G_{\langle q,r,s,p\rangle }(y)=1-\int_{F_{H_{q}\mid X_{q},Z_{q}}(0\mid
x,z)<v}\left[ 1-G_{s}\left( \frac{y}{Q_{H_{r}\mid X_{r},Z_{r}}(v\mid x,z)}%
,x,v\right) \right] dF_{X_{p},Z_{p},V_{p}}(x,z,v),  \label{eq1:g_qrst3}
\end{equation}%
where $v\mapsto Q_{H_{r}\mid X_{r},Z_{r}}(v\mid x,z)$\ is the quantile
(left-inverse) function of $h\mapsto F_{H_{r}\mid X_{r},Z_{r}}(h\mid x,z)$.
Equation (\ref{eq1:g_qrst3}) describes the mechanics for the construction of
the counterfactual distributions. Based on the numerator in the first
argument of the LDSF, it computes the hours of the individuals in the
population with composition as in year $p$ with the hours equation of year $%
r $ by keeping the individuals' ranks identical conditional on their
observed characteristics $X_{p}$ and $Z_{p}$. For example, an individual
working median annual hours in the subpopulation of individuals with the
same $X_{p} $ and $Z_{p}$ is assigned the median working hours in the same
subpopulation in year $r$. We do this for every individual and at each
quantile. An individual working in year $p$ \ may not work in year $r$ if
the corresponding quantile equals zero in year $r$. An interesting aspect of
(\ref{eq1:g_qrst3}) is the calculation of the area of integration. The
condition that defines this area compares the realization of the control
function in year $p$ with the fraction of individuals that do not work in
year $q,$ conditional on their values of $X_{p}$ and $Z_{p}$. If the value
of the control function is higher, then the individual would also work in
year $q$. Otherwise, that individual would not work.

We assume a random sample of size $n_{t}$, $\{(Y_{i,t},H_{i,t}\times
W_{i,t},H_{i,t},X_{i,t},Z_{i,t})\}_{i=1}^{n_{t}}$, from $(Y_{t},H_{t}\times
W_{t},H_{t},X_{t},Z_{t})$ for every year $t$, where $H_{t}\times W_{t}$
indicates that $W_{t}$ is only observed when $H_{t}>0$. We estimate the
control function via logistic distribution regression. For every observation
in the selected sample with $H_{i,t}>0$, we set:\label{page:estimation} 
\begin{equation*}
\widehat{V}_{i,t}=\Lambda (P_{i,t}^{\top }\widehat{\pi }_{t}(H_{i,t})),\quad
i=1,\ldots ,n_{t},
\end{equation*}%
where $\Lambda $ is the standard logistic distribution function, $%
P_{i,t}:=p(X_{i,t},Z_{i,t})$\ is a $d_{p}$-dimensional vector comprising of
transformations of $X_{i,t}$\ and $Z_{i,t}$, and $\widehat{\pi }_{t}(h)$\ is
the coefficient estimate in the logistic distribution regression of $\mathbf{%
1}(H_{i,t}\leq h)$ on $P_{i,t}$. With $\widehat{V}_{i,t}$ we estimate the
LDSF of hourly wages as $\widehat{G}_{t}(w,x,v)=\Lambda (m(x,v)^{\top }%
\widehat{\theta }_{t}(w)), $ where $m(x,v)$\ is a $d_{m}$-dimensional vector
comprising of transformations of $(x,v)$\ and the $\widehat{\theta }_{t}(w)$%
\ is the coefficient estimate in the logistic distribution regression of $%
\mathbf{1}(W_{i,t}\leq w)$ on $m(X_{i,t},\widehat{V}_{i,t})$.

With these different components we estimate $G_{\langle q,r,s,p\rangle }(y)$%
\ by the empirical counterpart of \eqref{eq1:g_qrst3}, namely: 
\begin{multline*}
G_{\langle q,r,s,p\rangle }(y)=1 \\
-\frac{1}{n_{s}}\sum_{i=1}^{n^{s}}\mathbf{1}\left( \Lambda (P_{i,s}^{\top }%
\widehat{\pi }_{q}(0))<\widehat{V}_{i,s}\right) \left[ 1-\widehat{G}%
_{p}\left( \frac{y}{\widehat{Q}_{H_{r}\mid X_{r},Z_{r}}(\widehat{V}%
_{i,s}\mid X_{i,s},Z_{i,s})},X_{i,s},Z_{i,s}\right) \right] ,
\end{multline*}%
where $\widehat{Q}_{H_{r}\mid X_{r},Z_{r}}(v\mid
X_{i,s},Z_{i,s})=\int_{0}^{\infty }1\{\Lambda (P_{i,s}^{\top }\widehat{\pi }%
_{r}(h))\leq v\}dh$\ is the generalized inverse of $h\mapsto \Lambda
(P_{i,s}^{\top }\widehat{\pi }_{r}(h))$.

The standard errors and confidence intervals of the components of the
decomposition can be calculated using weighted bootstrap. We report standard
errors for the decompositions in Appendix \ref{app:bootstraps} of Fern\'{a}%
ndez-Val et al. (2021b). The estimates are very precise due to the large
sample sizes.

\section{Empirical analysis: Preliminaries}

\label{section:preliminaries}

\subsection{Conditioning variables}

\label{ss:conditioning} Following Mulligan and Rubinstein (2008), we use 6
indicators to capture the highest level of education: (i) 0-8 years of
completed schooling, (ii) high school dropouts, (iii) high school graduates
or 12 years of schooling (including GEDs), (iv) some college, (v) college,
and (vi) advanced degree. Following Autor et al. (2016), we employ a
quadratic polynomial in potential experience fully interacted with the
education indicators. We include 4 indicators of marital status, and
indicators for African-American and Hispanic origin. We capture regional
variation with indicators for the 4 different regions defined in the CPS
(North East, North, West and South). The variables in $Z_{t}$ are the four
indicators capturing marital status interacted with an indicator for the
presence of children under the age of five. These household characteristics
are frequently used as exclusion restrictions (see, for example, Mulligan
and Rubinstein, 2008 for the CPS and Card and Hyslop, 2021 for the PSID) but
not without controversy.\label{page:exclusion_restriction} For example,
Blundell et al. (2007) argued that highly talented women delay childbirth
implying that conditional on age the family characteristics may be
correlated with unobserved ability. We explore the sensitivity to the
exclusion restrictions by including these variables in $X_{t}$ rather than $%
Z_{t}$ and find no notable change in the results. We accordingly employed
the above division of the variables into $X_{t}$ and $Z_{t}$.

\subsection{Treatment of bunching of the hours variable}

\label{ss:bunching}

Figure \ref{fig:hours} revealed substantial bunching of reported working
hours. This bunching, particularly at 2080 working hours per year, probably
reflects the convenience to respond 40 hours per week for 52 weeks per year.
However, it creates a difficulty in predicting an individual's working
hours. Accordingly, we modify our method as follows.

Suppose $H_{t}^{\ast }$\ is the actual number of the individuals' annual
hours and $H_{t}$\ is their observed annual hours. Hence, $H_{t}^{\ast }$\
is the left-hand side variable in (\ref{eq2}). Furthermore, assume $H_{t}=j%
\text{ if }h_{j-1}<H_{t}^{\ast }\leq h_{j}$. This implies that workers with
hours between cutoff values $h_{j-1}$ and $h_{j}$ report hours to be exactly
equal to $j$. The random variable $H_{t}$ will have a discrete distribution
although $H_{t}^{\ast }$ is continuous. From monotonicity $%
h_{j-1}<H_{t}^{\ast }\leq h_{j}$\ implies $V_{t}$\ is between the values $%
F_{H\mid X,Z}(h_{j-1}\mid x,z)$\ and $F_{H\mid X,Z}(h_{j}\mid x,z)$. Also, $%
V_{t}$ is uniformly distributed between these values although we are unable
to estimate the control function for the individual observations. A solution
is to estimate the values $F_{H\mid X,Z}(h_{j}\mid x,z)$ for every $j$ based
on the observed levels of working hours, and to randomly sample values
between the estimated $F_{H\mid X,Z}(h_{j-1}\mid x,z)$\ and $F_{H\mid
X,Z}(h_{j}\mid x,z)$ to obtain draws of $V_{t}$. We use these sampled values
rather than the estimated values based on $H_{t}$ to calculate the
counterfactual earnings distributions. 

\subsection{Choice of base year}

\label{ss:base_year}

The chosen base years are 1976 for females and 2010 for males. The base year
should have the lowest employment rates as this increases the likelihood
that the identification conditions are satisfied. The female employment rate
is at its lowest in 1976 and we assume that females with a certain
combination of $(x,z,v)$ working in 1976 will have a positive probability of
working in subsequent years. We choose 2010, the bottom of the financial
crisis, for males. As we employ different base years we can compare trends
but not hours and earnings across gender. Irrespective of the base year, the
results in Section \ref{sec:results} are relative to 1976. This has
implications, discussed below, for the decomposition figures.

\subsection{Estimation of the control function}

\label{ss:control_function_results}

We estimate the hours equation \eqref{eq2} separately by gender and survey
year and construct the control function $\widehat{V}_{t}$. We do not report
these results but highlight their implications for the earnings
decompositions. We estimate the hourly wage equation \eqref{eq1} for the
selected sample using $X_{t},$ $\widehat{V}_{t}$ and $\widehat{V}_{t}^{2},$
in addition to the product of $\widehat{V}_{t}$ with $X_{t}$, as
conditioning variables.

Recall what is required for identification and the correct interpretation of
our results.\label{page:support2} First are the support restrictions in
Proposition \ref{prop:id}. The most important is $\mathcal{XV}_{p}\cap 
\mathcal{XV}_{q}^{\ast }\subseteq \mathcal{XV}_{s}^{\ast }$ which is
satisfied whenever $\mathcal{XV}_{q}^{\ast }\subseteq \mathcal{XV}_{s}^{\ast
}$. This implies that the support of $V_{t}$ for any $X_{t}$ in the year $t$
should be at least as large as the base year support. To investigate this
note that for any $x\in \mathcal{X}_{t}^{\ast }$ the lower bound of $%
\mathcal{V}_{t}^{\ast }(x)$ equals $\min_{z\in \mathcal{Z}_{t}^{\ast
}(x)}F_{H_{t}|X_{t},Z_{t}}(0|x,z). $ As monotonicity imposes that the upper
bound always equals 1, Figure \ref{fig:check} of Fern\'andez-Val et al.
(2021b) reports the percentage of observations in year $t$ that have lower
bounds higher than the expected lower bounds in the base year. These are
zero by construction in 1976 for females and 2010 for males. For both males
and females the number of violations is very small.

Rank invariance is required to distinguish between extensive and intensive
effects.\label{page:rank_invariance2} A necessary condition is that the
conditional distributions of the control function do not differ between the
years. We investigate this through the stability of the estimated parameters
for the hours equation across years. We find the coefficients for two
important determinants of hours, education and age, to be stable over time.
This is supported by Figure \ref{fig:education_hours} of Fern\'{a}ndez-Val
et al. (2021b).


\section{Empirical Analysis: Results}

\label{sec:results}

\subsection{Earnings decompositions}

We discuss the results for males and females separately but present the
corresponding graphs together. Our primary focus is on the hours effects but
we also discuss the structural and composition effects. As we employ
different base years it is important to note the following. For females the
base year is 1976 so the interpretation is straightfoward. For males the
base year is 2010 and this imposes that the total effect, and each
individual component, is zero in that year. For presentational purposes we
shift figures illustrating the various effects to impose that they are all
set to zero in 1976.\footnote{%
For example, for the extensive effect, we use the following formula: $\left(
[Q_{\langle t,t,t,t\rangle }(\tau )-Q_{\langle 1976,1976,1976,1976\rangle
}(\tau )-Q_{\langle 2010,t,t,t\rangle }(\tau )+Q_{\langle 1976,t,t,t\rangle
}(\tau )]\right) /Q_{\langle 1976,1976,1976,1976\rangle }(\tau )$} This
facilitates an easier comparison across gender and a simpler representation
of changes over the sample period. One feature of the results which we do
not pursue but which merits noting is the impact of wages on hours. While we
do not impose sufficient structure to estimate labor supply elasticities we
do investigate the relationship between mean wages and the control function.\label{insert:elasticities}
The estimates are consistent with a positive relationship between wages and
annual hours and a positive wage elasticity for both males and females.

\subsubsection{Males}

\paragraph{Structural Wage Effects}

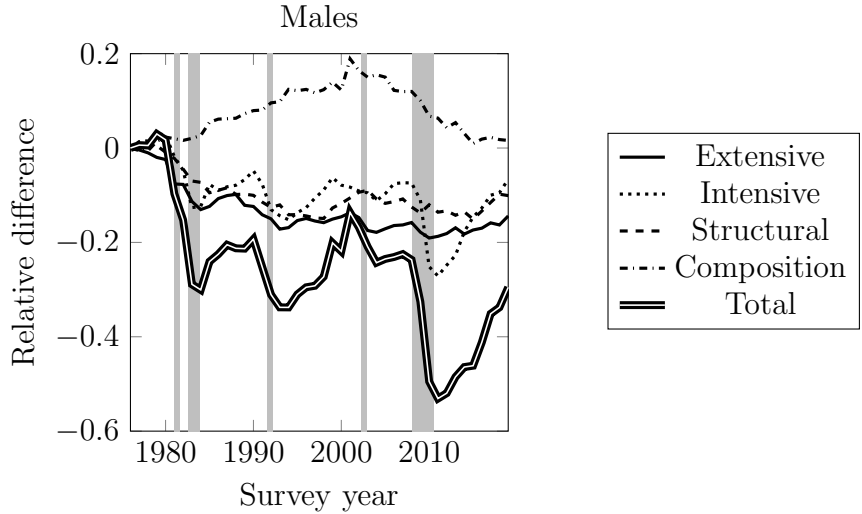
\begin{figure}[tbp]
\centering
\begin{tikzpicture}
	\begin{groupplot}
	[
	group style={%
		group size = 1 by 1,
		group name=plots,
		horizontal sep=0.5cm, 
		xlabels at=edge bottom,
		y descriptions at=edge left,
		ylabels at=edge left,
		x descriptions at=edge bottom
	},
	set layers,cell picture=true,
	width=0.4\textwidth,
	height=0.4\textwidth,
	legend columns=1,
	xlabel = Survey year,
	ylabel = Relative difference,
	ymin = -0.6,
	ymax = 0.2,
	xmin = 1976, 
	xmax = 2019,
	cycle list name=black white
	]
	\nextgroupplot[legend to name=grouplegend1, title = Males]
	\addplot[black, very thick] table[x = year, y = extensive]{results_quantile_male_25_2021.txt};
	\addlegendentry{Extensive}
	\addplot[black, dotted,  very thick] table[x = year, y = intensive]{results_quantile_male_25_2021.txt};
	\addlegendentry{Intensive}
	\addplot[black, dashed, very thick] table[x = year, y = structural]{results_quantile_male_25_2021.txt};	
	\addlegendentry{Structural}	
	\addplot[black, dashdotted, very thick] table[x = year, y = composition]{results_quantile_male_25_2021.txt};
	\addlegendentry{Composition}		
	\addplot[black, double, very thick] table[x = year, y = total]{results_quantile_male_25_2021.txt};			
	\addlegendentry{Total}		
	\addplot[domain=1981.0:1981.6, name path = C]{25};
	\addplot[domain=1981.0:1981.6, name path = D]{-5};      
	\addplot[domain=1982.6:1983.92, name path = E]{25};
	\addplot[domain=1982.6:1983.92, name path = F]{-5};      		
	\addplot[domain=1991.6:1992.25, name path = G]{25};
	\addplot[domain=1991.6:1992.25, name path = H]{-5};   	 
	\addplot[domain=2002.25:2002.92, name path = I]{25};
	\addplot[domain=2002.25:2002.92, name path = J]{-5};   			
	\addplot[domain=2008.09:2010.5, name path = K]{25};
	\addplot[domain=2008.09:2010.5, name path = L]{-5};   		
	\addplot[lightgray] fill between[of=C and D];
	\addplot[lightgray] fill between[of=E and F];
	\addplot[lightgray] fill between[of=G and H];           
	\addplot[lightgray] fill between[of=I and J];
	\addplot[lightgray] fill between[of=K and L]; 	
	\end{groupplot}
	\node at (plots c1r1.east) [inner sep=10pt,anchor=north, xshift= 3cm,yshift=10ex] {\ref{grouplegend1}};  
	\end{tikzpicture}
\caption{Decomposition of the relative changes in the lower quartile of
annual earnings with respect to 1976.}
\label{fig:decomp_q1}
\end{figure}

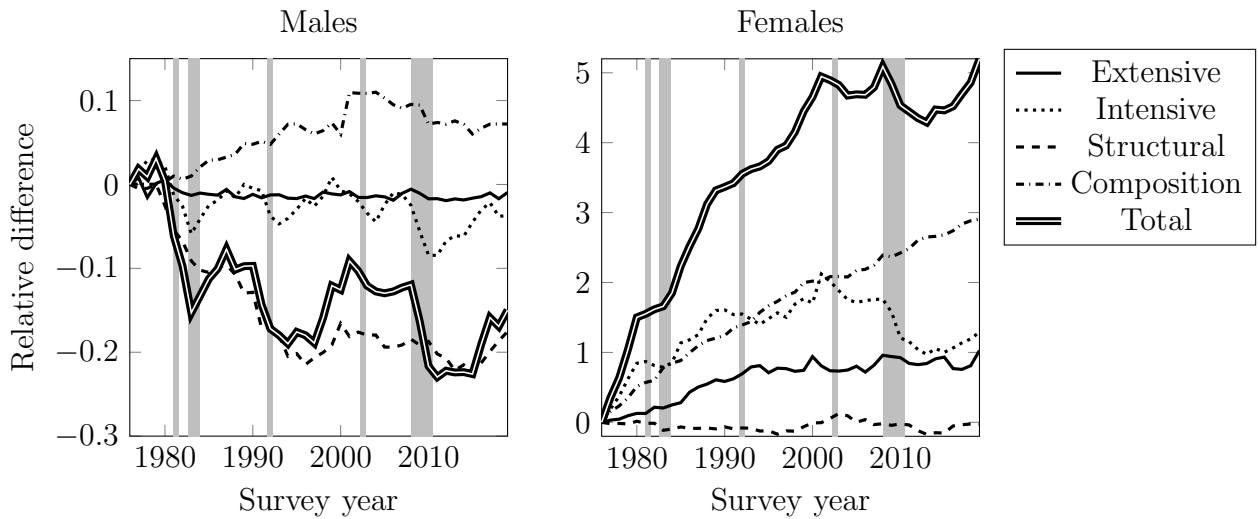
\begin{figure}[tbp]
\centering
\begin{tikzpicture}
	\begin{groupplot}
	[
	group style={%
		group size = 2 by 1,
		group name=plots
		, horizontal sep=1.25cm, 
		xlabels at=edge bottom,
		ylabels at=edge left,
		x descriptions at=edge bottom
	},
	set layers,cell picture=true,
	width=0.4\textwidth,
	height=0.4\textwidth,
	legend columns=1,
	xlabel = Survey year,
	ylabel = Relative difference,
	ymin = -0.2,
	ymax = 5.2,
	xmin = 1976, 
	xmax = 2019,
	cycle list name=black white
	]
	\nextgroupplot[legend to name=grouplegend1, title = Males, ymin = -0.3, ymax = 0.15]
	\addplot[black, very thick] table[x = year, y = extensive]{results_quantile_male_50_2021.txt};
	\addlegendentry{Extensive}
	\addplot[black, dotted,  very thick] table[x = year, y = intensive]{results_quantile_male_50_2021.txt};
	\addlegendentry{Intensive}
	\addplot[black, dashed, very thick] table[x = year, y = structural]{results_quantile_male_50_2021.txt};
	\addlegendentry{Structural}	
	\addplot[black, dashdotted, very thick] table[x = year, y = composition]{results_quantile_male_50_2021.txt};
	\addlegendentry{Composition}		
	\addplot[black, double, very thick] table[x = year, y = total]{results_quantile_male_50_2021.txt};		
	\addlegendentry{Total}		
	\addplot[domain=1981.0:1981.6, name path = C]{25};
	\addplot[domain=1981.0:1981.6, name path = D]{-5};      
	\addplot[domain=1982.6:1983.92, name path = E]{25};
	\addplot[domain=1982.6:1983.92, name path = F]{-5};      		
	\addplot[domain=1991.6:1992.25, name path = G]{25};
	\addplot[domain=1991.6:1992.25, name path = H]{-5};   	 
	\addplot[domain=2002.25:2002.92, name path = I]{25};
	\addplot[domain=2002.25:2002.92, name path = J]{-5};   			
	\addplot[domain=2008.09:2010.5, name path = K]{25};
	\addplot[domain=2008.09:2010.5, name path = L]{-5};   		
	\addplot[lightgray] fill between[of=C and D];
	\addplot[lightgray] fill between[of=E and F];
	\addplot[lightgray] fill between[of=G and H];           
	\addplot[lightgray] fill between[of=I and J];
	\addplot[lightgray] fill between[of=K and L]; 		
	\nextgroupplot[title = Females]
	\addplot[black, very thick] table[x = year, y = extensive]{results_quantile_female_50_2021.txt};
	\addplot[black, dotted,  very thick] table[x = year, y = intensive]{results_quantile_female_50_2021.txt};
	\addplot[black, dashed, very thick] table[x = year, y = structural]{results_quantile_female_50_2021.txt};
	\addplot[black, dashdotted, very thick] table[x = year, y = composition]{results_quantile_female_50_2021.txt};	
	\addplot[black, double, very thick] table[x = year, y = total]{results_quantile_female_50_2021.txt};		
	\addplot[domain=1981.0:1981.6, name path = C]{25};
	\addplot[domain=1981.0:1981.6, name path = D]{-5};      
	\addplot[domain=1982.6:1983.92, name path = E]{25};
	\addplot[domain=1982.6:1983.92, name path = F]{-5};      		
	\addplot[domain=1991.6:1992.25, name path = G]{25};
	\addplot[domain=1991.6:1992.25, name path = H]{-5};   	 
	\addplot[domain=2002.25:2002.92, name path = I]{25};
	\addplot[domain=2002.25:2002.92, name path = J]{-5};   			
	\addplot[domain=2008.09:2010.5, name path = K]{25};
	\addplot[domain=2008.09:2010.5, name path = L]{-5};   		
	\addplot[lightgray] fill between[of=C and D];
	\addplot[lightgray] fill between[of=E and F];
	\addplot[lightgray] fill between[of=G and H];           
	\addplot[lightgray] fill between[of=I and J];
	\addplot[lightgray] fill between[of=K and L]; 
	\end{groupplot}
	\node at (plots c2r1.east) [inner sep=10pt,anchor=north, xshift = 2cm, yshift=3cm] {\ref{grouplegend1}};  
	\end{tikzpicture}
\caption{Decomposition of the relative changes in median annual earnings
with respect to 1976.}
\label{fig:decomp_q2}
\end{figure}

\begin{figure}[tbp]
\centering
\begin{tikzpicture}
	\begin{groupplot}
	[
	group style={%
		group size = 2 by 1,
		group name=plots
		, horizontal sep=1.25cm, 
		xlabels at=edge bottom,
		ylabels at=edge left,
		x descriptions at=edge bottom
	},
	set layers,cell picture=true,
	width=0.4\textwidth,
	height=0.4\textwidth,
	legend columns=1,
	xlabel = Survey year,
	ylabel = Relative difference,
	ymin = -0.2,
	ymax = 0.9,
	xmin = 1976, 
	xmax = 2019,
	cycle list name=black white
	]
	\nextgroupplot[legend to name=grouplegend1, title = Males]
	\addplot[black, very thick] table[x = year, y = extensive]{results_quantile_male_75_2021.txt};
	\addlegendentry{Extensive}
	\addplot[black, dotted,  very thick] table[x = year, y = intensive]{results_quantile_male_75_2021.txt};
	\addlegendentry{Intensive}
	\addplot[black, dashed, very thick] table[x = year, y = structural]{results_quantile_male_75_2021.txt};		
	\addlegendentry{Structural}	
	\addplot[black, dashdotted, very thick] table[x = year, y = composition]{results_quantile_male_75_2021.txt};
	\addlegendentry{Composition}		
	\addplot[black, double, very thick] table[x = year, y = total]{results_quantile_male_75_2021.txt};				
	\addlegendentry{Total}		
	\addplot[domain=1981.0:1981.6, name path = C]{25};
	\addplot[domain=1981.0:1981.6, name path = D]{-5};      
	\addplot[domain=1982.6:1983.92, name path = E]{25};
	\addplot[domain=1982.6:1983.92, name path = F]{-5};      		
	\addplot[domain=1991.6:1992.25, name path = G]{25};
	\addplot[domain=1991.6:1992.25, name path = H]{-5};   	 
	\addplot[domain=2002.25:2002.92, name path = I]{25};
	\addplot[domain=2002.25:2002.92, name path = J]{-5};   			
	\addplot[domain=2008.09:2010.5, name path = K]{25};
	\addplot[domain=2008.09:2010.5, name path = L]{-5};   		
	\addplot[lightgray] fill between[of=C and D];
	\addplot[lightgray] fill between[of=E and F];
	\addplot[lightgray] fill between[of=G and H];           
	\addplot[lightgray] fill between[of=I and J];
	\addplot[lightgray] fill between[of=K and L]; 			
	\nextgroupplot[title = Females]
	\addplot[black, very thick] table[x = year, y = extensive]{results_quantile_female_75_2021.txt};
	\addplot[black, dotted,  very thick] table[x = year, y = intensive]{results_quantile_female_75_2021.txt};
	\addplot[black, dashed, very thick] table[x = year, y = structural]{results_quantile_female_75_2021.txt};	
	\addplot[black, dashdotted, very thick] table[x = year, y = composition]{results_quantile_female_75_2021.txt};
	\addplot[black, double, very thick] table[x = year, y = total]{results_quantile_female_75_2021.txt};			
	\addplot[domain=1981.0:1981.6, name path = C]{25};
	\addplot[domain=1981.0:1981.6, name path = D]{-5};      
	\addplot[domain=1982.6:1983.92, name path = E]{25};
	\addplot[domain=1982.6:1983.92, name path = F]{-5};      		
	\addplot[domain=1991.6:1992.25, name path = G]{25};
	\addplot[domain=1991.6:1992.25, name path = H]{-5};   	 
	\addplot[domain=2002.25:2002.92, name path = I]{25};
	\addplot[domain=2002.25:2002.92, name path = J]{-5};   			
	\addplot[domain=2008.09:2010.5, name path = K]{25};
	\addplot[domain=2008.09:2010.5, name path = L]{-5};   		
	\addplot[lightgray] fill between[of=C and D];
	\addplot[lightgray] fill between[of=E and F];
	\addplot[lightgray] fill between[of=G and H];           
	\addplot[lightgray] fill between[of=I and J];
	\addplot[lightgray] fill between[of=K and L]; 
	\end{groupplot}
	\node at (plots c2r1.east) [inner sep=10pt,anchor=north, xshift = 2cm, yshift=3cm] {\ref{grouplegend1}};  
	\end{tikzpicture}
\caption{Decomposition of the relative changes in the upper quartile of
annual earnings with respect to 1976.}
\label{fig:decomp_q3b}
\end{figure}
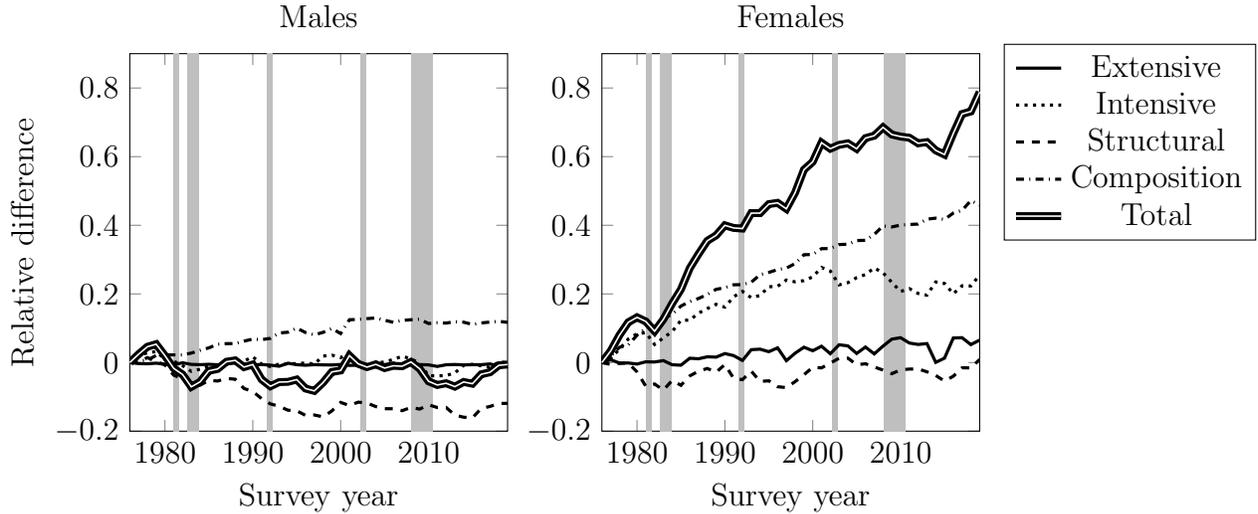

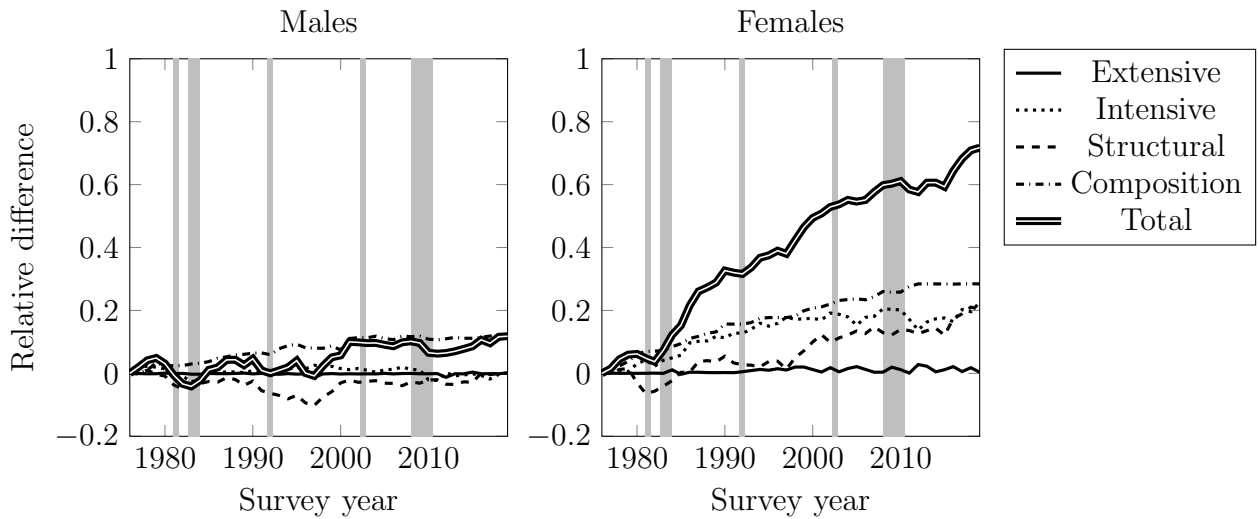
\begin{figure}[tbp]
\begin{tikzpicture}
	\begin{groupplot}
	[
	group style={%
		group size = 2 by 1,
		group name=plots
		, horizontal sep=1.25cm, 
		xlabels at=edge bottom,
		ylabels at=edge left,
		x descriptions at=edge bottom
	},
	set layers,cell picture=true,
	width=0.4\textwidth,
	height=0.4\textwidth,
	legend columns=1,
	xlabel = Survey year,
	ylabel = Relative difference,
	ymin = -0.2,
	ymax = 1,
	xmin = 1976, 
	xmax = 2019,
	cycle list name=black white
	]
	\nextgroupplot[legend to name=grouplegend1, title = Males]
	\addplot[black, very thick] table[x = year, y = extensive]{results_quantile_male_90_2021.txt};
	\addlegendentry{Extensive}
	\addplot[black, dotted,  very thick] table[x = year, y = intensive]{results_quantile_male_90_2021.txt};
	\addlegendentry{Intensive}
	\addplot[black, dashed, very thick] table[x = year, y = structural]{results_quantile_male_90_2021.txt};		
	\addlegendentry{Structural}	
	\addplot[black, dashdotted, very thick] table[x = year, y = composition]{results_quantile_male_90_2021.txt};
	\addlegendentry{Composition}		
	\addplot[black, double, very thick] table[x = year, y = total]{results_quantile_male_90_2021.txt};				
	\addlegendentry{Total}		
	\addplot[domain=1981.0:1981.6, name path = C]{25};
	\addplot[domain=1981.0:1981.6, name path = D]{-5};      
	\addplot[domain=1982.6:1983.92, name path = E]{25};
	\addplot[domain=1982.6:1983.92, name path = F]{-5};      		
	\addplot[domain=1991.6:1992.25, name path = G]{25};
	\addplot[domain=1991.6:1992.25, name path = H]{-5};   	 
	\addplot[domain=2002.25:2002.92, name path = I]{25};
	\addplot[domain=2002.25:2002.92, name path = J]{-5};   			
	\addplot[domain=2008.09:2010.5, name path = K]{25};
	\addplot[domain=2008.09:2010.5, name path = L]{-5};   		
	\addplot[lightgray] fill between[of=C and D];
	\addplot[lightgray] fill between[of=E and F];
	\addplot[lightgray] fill between[of=G and H];           
	\addplot[lightgray] fill between[of=I and J];
	\addplot[lightgray] fill between[of=K and L]; 			
	\nextgroupplot[title = Females]
	\addplot[black, very thick] table[x = year, y = extensive]{results_quantile_female_90_2021.txt};
	\addplot[black, dotted,  very thick] table[x = year, y = intensive]{results_quantile_female_90_2021.txt};
	\addplot[black, dashed, very thick] table[x = year, y = structural]{results_quantile_female_90_2021.txt};	
	\addplot[black, dashdotted, very thick] table[x = year, y = composition]{results_quantile_female_90_2021.txt};
	\addplot[black, double, very thick] table[x = year, y = total]{results_quantile_female_90_2021.txt};			
	\addplot[domain=1981.0:1981.6, name path = C]{25};
	\addplot[domain=1981.0:1981.6, name path = D]{-5};      
	\addplot[domain=1982.6:1983.92, name path = E]{25};
	\addplot[domain=1982.6:1983.92, name path = F]{-5};      		
	\addplot[domain=1991.6:1992.25, name path = G]{25};
	\addplot[domain=1991.6:1992.25, name path = H]{-5};   	 
	\addplot[domain=2002.25:2002.92, name path = I]{25};
	\addplot[domain=2002.25:2002.92, name path = J]{-5};   			
	\addplot[domain=2008.09:2010.5, name path = K]{25};
	\addplot[domain=2008.09:2010.5, name path = L]{-5};   		
	\addplot[lightgray] fill between[of=C and D];
	\addplot[lightgray] fill between[of=E and F];
	\addplot[lightgray] fill between[of=G and H];           
	\addplot[lightgray] fill between[of=I and J];
	\addplot[lightgray] fill between[of=K and L]; 
	\end{groupplot}
	\node at (plots c2r1.east) [inner sep=10pt,anchor=north, xshift = 2cm, yshift=3cm] {\ref{grouplegend1}};  
	\end{tikzpicture}
\caption{Decomposition of the relative changes in the upper decile of annual
earnings with respect to 1976.}
\label{fig:decomp_q9b}
\end{figure}

Figure \ref{fig:decomp_mean} shows changes in mean earnings are generally
driven by the large and negative structural wage effects. This partially
reflects the lower wage profiles of younger generations (see, for example,
Kambourov and Minovskii, 2009). The structural wage effect increases from
the mid 1990s but remains negative until the end of the sample period.
Figure~\ref{fig:decomp_q1} examines the lower quartile of annual earnings
noting this corresponds to the first decile of earners. The structural wage
effect is an important component of the decreases and exacerbates the
earnings reductions from nonemployment during cyclical downturns discussed
below. There are negative structural wage effects throughout the entire
sample period at this quantile. Figure~\ref{fig:decomp_q2} reveals that
negative structural wage effects are also the driving influence in the
evolution in median earnings. This is consistent with existing work on the
evaluation of workers' characteristics in the middle of the male wage
distribution. The large negative structural effects at Q3 and the small
effects at D9, shown in Figures \ref{fig:decomp_q3b} and \ref{fig:decomp_q9b}%
, are also consistent with existing empirical work.\vspace{-0.4cm}

\pgfplotstableread{results_mean_male_1_2020_2.txt}{\resultsa} %
\pgfplotstableread{results_mean_male_2_2020_2.txt}{\resultsaa} %
\pgfplotstableread{results_mean_male_3_2020_2.txt}{\resultsab} %
\pgfplotstableread{results_mean_male_4_2020_2.txt}{\resultsac} %
\pgfplotstableread{results_mean_male_5_2020_2.txt}{\resultsad}

\pgfplotstableread{results_mean_female_1_2020_2.txt}{\resultsb} %
\pgfplotstableread{results_mean_female_2_2020_2.txt}{\resultsba} %
\pgfplotstableread{results_mean_female_3_2020_2.txt}{\resultsbb} %
\pgfplotstableread{results_mean_female_4_2020_2.txt}{\resultsbc} %
\pgfplotstableread{results_mean_female_5_2020_2.txt}{\resultsbd}

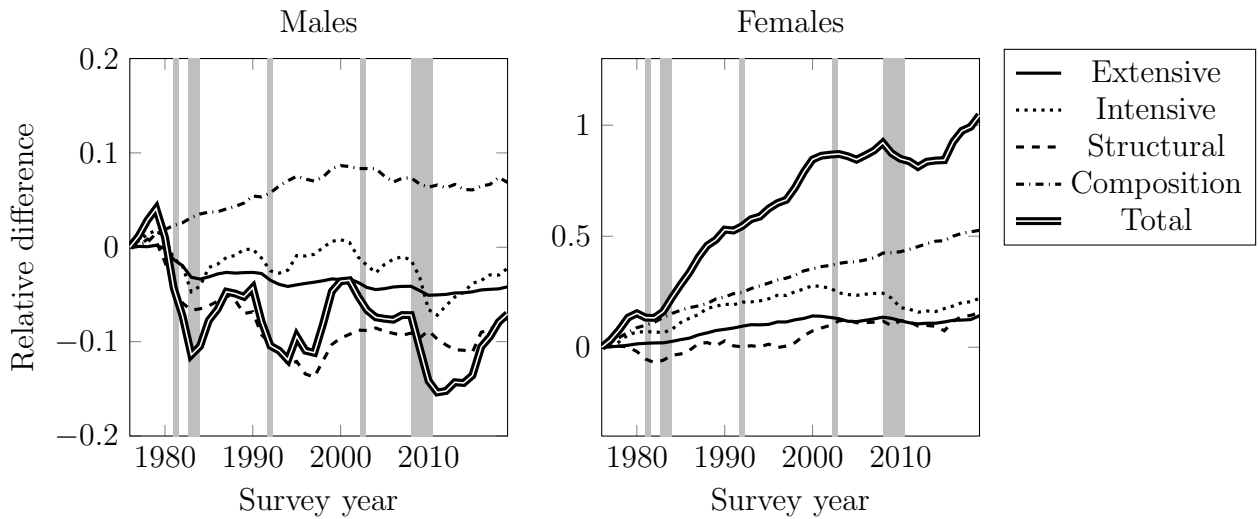
\begin{figure}[tbp]
\centering
\begin{tikzpicture}
	\begin{groupplot}
	[
	group style={%
		group size = 2 by 1,
		group name=plots
		, horizontal sep=1.25cm, 
		xlabels at=edge bottom,
		ylabels at=edge left,
		x descriptions at=edge bottom
	},
	set layers,cell picture=true,
	width=0.4\textwidth,
	height=0.4\textwidth,
	legend columns=1,
	xlabel = Survey year,
	ylabel = Relative difference,
	ymin = -0.4,
	ymax = 1.3,
	xmin = 1976, 
	xmax = 2019,
	cycle list name=black white
	]
	\nextgroupplot[legend to name=grouplegend1, title = Males, ymin = -0.2, ymax = 0.2]
	\addplot[black, very thick] table{\resultsa};
	\addlegendentry{Extensive}
	\addplot[black, dotted,  very thick] table{\resultsaa};
	\addlegendentry{Intensive}
	\addplot[black, dashed, very thick] table{\resultsab};
	\addlegendentry{Structural}	
	\addplot[black, dashdotted, very thick] table{\resultsac};
	\addlegendentry{Composition}		
	\addplot[black, double, very thick] table{\resultsad};		
	\addlegendentry{Total}		
	\addplot[domain=1981.0:1981.6, name path = C]{25};
	\addplot[domain=1981.0:1981.6, name path = D]{-5};      
	\addplot[domain=1982.6:1983.92, name path = E]{25};
	\addplot[domain=1982.6:1983.92, name path = F]{-5};      		
	\addplot[domain=1991.6:1992.25, name path = G]{25};
	\addplot[domain=1991.6:1992.25, name path = H]{-5};   	 
	\addplot[domain=2002.25:2002.92, name path = I]{25};
	\addplot[domain=2002.25:2002.92, name path = J]{-5};   			
	\addplot[domain=2008.09:2010.5, name path = K]{25};
	\addplot[domain=2008.09:2010.5, name path = L]{-5};   		
	\addplot[lightgray] fill between[of=C and D];
	\addplot[lightgray] fill between[of=E and F];
	\addplot[lightgray] fill between[of=G and H];           
	\addplot[lightgray] fill between[of=I and J];
	\addplot[lightgray] fill between[of=K and L]; 		
	\nextgroupplot[title = Females]
	\addplot[black, very thick] table{\resultsb};
	\addplot[black, dotted,  very thick] table{\resultsba};
	\addplot[black, dashed, very thick] table{\resultsbb};
	\addplot[black, dashdotted, very thick] table{\resultsbc};
	\addplot[black, double, very thick] table{\resultsbd};			
	\addplot[domain=1981.0:1981.6, name path = C]{25};
	\addplot[domain=1981.0:1981.6, name path = D]{-5};      
	\addplot[domain=1982.6:1983.92, name path = E]{25};
	\addplot[domain=1982.6:1983.92, name path = F]{-5};      		
	\addplot[domain=1991.6:1992.25, name path = G]{25};
	\addplot[domain=1991.6:1992.25, name path = H]{-5};   	 
	\addplot[domain=2002.25:2002.92, name path = I]{25};
	\addplot[domain=2002.25:2002.92, name path = J]{-5};   			
	\addplot[domain=2008.09:2010.5, name path = K]{25};
	\addplot[domain=2008.09:2010.5, name path = L]{-5};   		
	\addplot[lightgray] fill between[of=C and D];
	\addplot[lightgray] fill between[of=E and F];
	\addplot[lightgray] fill between[of=G and H];           
	\addplot[lightgray] fill between[of=I and J];
	\addplot[lightgray] fill between[of=K and L]; 
	\end{groupplot}
	\node at (plots c2r1.east) [inner sep=10pt,anchor=north, xshift = 2cm, yshift=3cm] {\ref{grouplegend1}};  
	\end{tikzpicture}
\caption{Decomposition of the changes in mean annual earnings relative to
1976.}
\label{fig:decomp_mean}
\end{figure}

\paragraph{Composition Effects}

As there is no evidence of composition effects in the male hours
distribution the reported effects are operating through their impact on
hourly wages. At the mean and the considered quantiles the composition
effect is positive and frequently large. This reflects the increased
educational attainment of males. This is consistent with earlier studies
examining the role of composition effects on the wage distribution.\vspace{%
-0.4cm}

\paragraph{Hours Effects}

The hours effects at mean earnings are negative and economically important.%
\footnote{%
The hours effects are constructed to be positive relative to the base year.
For males we choose 2010 as the base year but normalize the total effect to
be zero in 1976 for easy comparisons across gender. This allows the observed
hours effects to be negative.} The extensive margin is generally the more
important although the intensive effect is frequently of a similar
magnitude. The intensive effect appears to be particularly important in
cyclical downturns. Figure \ref{fig:decomp_q1} indicates that the
substantial reductions in earnings at Q1 in recessionary periods reflect
movements at both the extensive and intensive margins. The biggest effects
are at the intensive margin during the recessionary periods although
movements at the extensive margin also substantially decrease earnings. The
very large intensive margin effect during the Great Recession probably
includes spells of unemployment in reduced weeks worked.\label%
{insert:great_recession} The total hours effect is large and economically
important and represents the largest component of the change in earnings.
The behavior of male earnings at this quantile is one of this paper's more
important findings and supports the following conclusions.\label%
{page:economic_insight copy(1)} First, while previous studies have
documented the cyclical sensitivity of wages at this quantile (see, for
example Elsby et al, 2016), we find that these wage decreases are
accompanied by non employment and large reductions in hours worked. Hoynes
et al. (2012) report that the unemployment and employment levels of lowly
educated males of Black and Hispanic ethnicity display large cyclical
patterns.\footnote{%
We conducted our empirical analysis separately by racial subgroups and found
that the earnings behavior of Hispanics and Blacks displayed greater
cyclicality than Whites. However, this appears to reflect their location in
the earnings distribution rather than heterogeneous impacts by race.} Our
findings support this result given these are frequently the lowest wage
workers. Similar to Hoynes et al. (2012), we find evidence of this
relationship in every cyclical downturn with a substantially greater impact
in the Great Recession. Second, we find that the intensive hours effect
recovers quickly after each recession. This is consistent with Altonji et
al. (2013) who find that unemployment spells are associated with large
reductions in working hours in the first year following unemployment
although wage decreases do not persist.\label{page:altonji copy(1)} Third,
although the intensive hours effects are highly cyclical, the structural
wage effect, especially during the Great Recession, is less so. That is, the
large earnings reduction experienced at this quantile during this recession
are due to decreases in annual hours. This supports the findings of Altonji
et al. (2013), Beaudry et al. (2016) and Valetta (2017) who find that highly
educated workers may have substituted workers at the bottom of the
distribution. This reduced the employment levels of these displaced workers
but did not greatly affect the wages of those employed.\label{page:beaudry2
copy(1)} It is also consistent with Elsby et al. (2016) who find that wage
declines during the recessions in the early 1980s were larger than those of
the Great Recession. Finally, this reduction in hours exacerbates the
inequality generated by wage differences.

Figure~\ref{fig:decomp_q2} presents the decomposition for male median annual
earnings and reveals that the intensive hours effects are relatively small
with the exception of notable contributions during recessionary periods.
There is no evidence of extensive hours effects recalling part of the
intensive hours effect reflects an increase in the number of weeks of
positive hours. However, a comparison with Figure~\ref{fig:decomp_q1}
reveals the large hours effects reinforcing the negative structural wage
effects at Q1 are smaller at Q2. This suggests that the reductions in hours
are disproportionately borne by those at the lower part of the wage and
earnings distributions. The decompositions for Q3 and D9, shown in Figures %
\ref{fig:decomp_q3b} and \ref{fig:decomp_q9b} respectively, support this
view. At neither quantile is there evidence of substantial hours effects. Q3
of male annual earnings is reasonably flat and the substantial increase at
D9 reflects the growth in hourly wage rates at this point of the wage
distribution.

\subsubsection{Females}

\paragraph{Structural Wage Effects}

Figure \ref{fig:decomp_mean} reveals that\ while mean female earnings
increased by over 100\% a relatively small fraction is due to the structural
wage effect. Previous work has established that the structural wage effect
increases female hourly wages at the upper end of the female wage
distribution but reduces wages at lower quantiles. The positive effects here
appear to dominate towards the end of the time period. Figure~\ref%
{fig:decomp_q2} also reveals that virtually none of the dramatic increase in
median earnings over the sample period can be attributed to the structural
wage effect. In fact, for a substantial part of the sample period the
structural wage effect is negative. A similar result is found for Q3. At D9
the structural wage effect is initially negative but grows through the
sample period and is large and positive by 2019.\vspace{-0.4cm}

\paragraph{Composition Effects}

The composition effect, operating both through hours and wages, is the
driving influence behind the growth in mean earnings and accounts for over
half of the increase by 2019. It reflects the large increase in female
educational attainment which increased both annual hours and wage rates.
Mean annual hours increase by around 50\% over the sample period and nearly
half of this is due to the composition effect.\footnote{%
Female mean annual hours worked increase from 1515 to 1867 over the sample
period.} The composition effect at Q2 is large and operates through both
wages and hours.\footnote{%
Those with median earnings are not necessarily working median hours.
However, it is worth noting that median hours increase by 150 percent and 50
percent of this change is due to composition effects.} The composition
effect is known to increase wages at all points of the female wage
distribution (see, for example, FPVV). However it also captures the impact
of higher education levels on annual hours. It contributes more than half of
the observed growth at Q2. A similar result is observed at Q3 although the
growth in earnings is less dramatic. At D9 the contribution falls to
slightly less than 50\% of the total growth in earnings.\footnote{%
The decompositions of hours distributions revealed no composition effects
above the median. While we continue to acknowledge that observations at the
upper quantiles of earnings are not necessarily the same as those in upper
quantiles of hours distribution it is very likely that there is substantial
overlap.}\vspace{-0.4cm}

\paragraph{Hours Effects}

For mean earnings the intensive hours effect is substantial and reflects the
drastic shift in the females hours distribution. This has increased females
earnings by around 25\%. The extensive hours effect is slightly smaller and
is similar in magnitude to the structural wage effect by 2019. This differs
somewhat from the conclusion of Card and Hyslop (2021) who find large
extensive effects on inequality.\label{page:card_hyslop copy(4)} We pursue
the causes of this relatively modest increase below.

Movement on the intensive margin at the median is producing an increase of
around 125\% and this represents around a quarter of the total increase.
This reflects the large increase in annual hours for those at the lower part
of the hours distribution. The contribution of movements at the extensive
margin is surprisingly small and probably captures our assumption that those
who entered the labor market are those with the lowest number of annual
hours.

The estimated hours effects are interesting from a number of perspectives.%
\label{page:economic_insight_1 copy(1)} The intensive margin is generally
increasing from 1976 to 2000 before leveling off and decreasing for the
remainder of the sample period. The large cyclical effects operating through
the intensive margin for males are much smaller for females. While some of
the recessionary periods witness declines, the impact on median earnings is
relatively small. The exception is the Great Recession which has a
remarkable decline via the intensive hours effect. This supports Hoynes et
al. (2012) who conclude that female minority groups suffered substantially
during this period. An even more striking result is the smaller extensive
margin effect although by\ 2019 it is similar in magnitude to the intensive
margin effect and the total hours effect is around 200\%. This represents
approximately 40\% of the total change in median earnings. Recall that, due
to the rank invariance assumption, the decomposition exercise compares
different years to the base year and this deletes the individuals with very
few annual hours, and very low annual earnings, in calculating the extensive
hours effects. Although there are points in the distribution where this will
make a difference, generally only the bottom part of the hours distribution
is relevant. The evidence here suggests that these deletions are of
secondary importance in comparison to the changing hours distribution of
those working. However, this requires the rank invariance assumption and we
further investigate this below.

While the hours effects decrease as we examine the higher levels of
earnings, they remain important at both Q3 and D9. At each of these
quantiles the increase in earnings is substantial and a relatively large
contribution comes from movements on the intensive margin. This illustrates
how drastic the change in the female hours distributions has been. At both
Q3 and D9 the movements at the intensive margin explain around one quarter
of the total increases in earnings. The movements in the extensive margin
are not substantial at either quantile.

\pgfplotstableread{results_quantile_male_1_75_2020.txt}{\resultsa} %
\pgfplotstableread{results_quantile_male_2_75_2020.txt}{\resultsaa} %
\pgfplotstableread{results_quantile_male_3_75_2020.txt}{\resultsab} %
\pgfplotstableread{results_quantile_male_4_75_2020.txt}{\resultsac} %
\pgfplotstableread{results_quantile_male_5_75_2020.txt}{\resultsad} %

\pgfplotstableread{results_quantile_female_1_75_2020.txt}{\resultsb} %
\pgfplotstableread{results_quantile_female_2_75_2020.txt}{\resultsba} %
\pgfplotstableread{results_quantile_female_3_75_2020.txt}{\resultsbb} %
\pgfplotstableread{results_quantile_female_4_75_2020.txt}{\resultsbc} %
\pgfplotstableread{results_quantile_female_5_75_2020.txt}{\resultsbd} %
\pgfplotstableread{results_quantile_female_6_75.txt}{\resultsbe}

\subsubsection{Intensive or Extensive Margin?}

\label{sss:intensive_extensive}

The small contribution of the extensive effects seems puzzling.\label%
{page:extensive_hours} One reason is that the component of the extensive
margin capturing the impact of increased education on employment rates is
attributed here to the composition effect. A second cause is the
monotonicity assumption. Recall that the extensive effect is calculated by
setting the working hours of the marginal workers equal to zero and
evaluating the impact of such an exercise on the earnings distribution.
Monotonicity imposes that the marginal workers are those with the lowest
working hours. As these workers have generally low earnings, the impact of
their exclusion will be small.\label{insert:explain} This is especially true
if the distribution of annual hours features a large fraction of workers
with low annual hours. Another possibility is our use of a year as the unit
of measurement. This implies that short spells of unemployment are captured
in the intensive rather than the extensive hours effect. It is also
responsible for the small extensive hours effects for males during the Great
Recession. \label{insert:extensive_intensive} Other empirical studies have
experienced related problems associated with short spells of unemployment
(Altonji, et al., 2013).\label{insert:altonji1} The use of cross sectional
data requires a definition of movement on the extensive margin and this has
implications for the intensive margin. This is also noted by Blundell et al.
(2013).\label{insert:blundell}

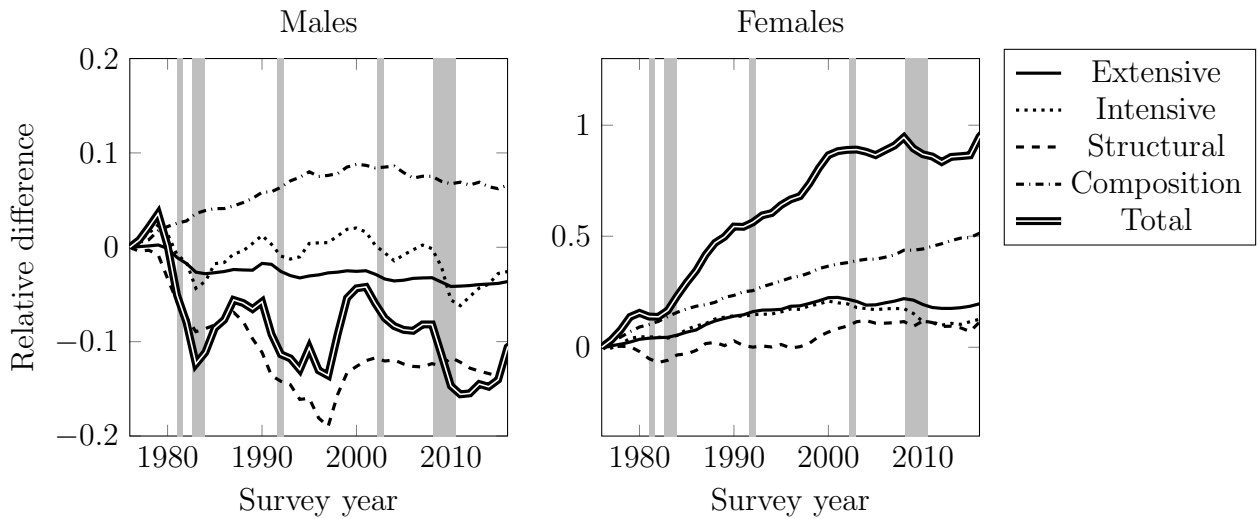
\begin{figure}[tbp]
\centering
\begin{tikzpicture}
	\begin{groupplot}
	[
	group style={%
		group size = 2 by 1,
		group name=plots
		, horizontal sep=1.25cm, 
		xlabels at=edge bottom,
		ylabels at=edge left,
		x descriptions at=edge bottom
	},
	set layers,cell picture=true,
	width=0.4\textwidth,
	height=0.4\textwidth,
	legend columns=1,
	xlabel = Survey year,
	ylabel = Relative difference,
	ymin = -0.4,
	ymax = 1.3,
	xmin = 1976, 
	xmax = 2016,
	cycle list name=black white
	]
	\nextgroupplot[legend to name=grouplegend1, title = Males, ymin = -0.2, ymax = 0.2]
	\addplot[black, very thick] table[x = year, y = extensive]{results_mean_male_2021_500.txt};
\addlegendentry{Extensive}
\addplot[black, dotted,  very thick] table[x = year, y = intensive]{results_mean_male_2021_500.txt};
\addlegendentry{Intensive}
\addplot[black, dashed, very thick] table[x = year, y = structural]{results_mean_male_2021_500.txt};		
\addlegendentry{Structural}	
\addplot[black, dashdotted, very thick] table[x = year, y = composition]{results_mean_male_2021_500.txt};
\addlegendentry{Composition}		
\addplot[black, double, very thick] table[x = year, y = total]{results_mean_male_2021_500.txt};				
\addlegendentry{Total}		
	\addplot[domain=1981.0:1981.6, name path = C]{25};
	\addplot[domain=1981.0:1981.6, name path = D]{-5};      
	\addplot[domain=1982.6:1983.92, name path = E]{25};
	\addplot[domain=1982.6:1983.92, name path = F]{-5};      		
	\addplot[domain=1991.6:1992.25, name path = G]{25};
	\addplot[domain=1991.6:1992.25, name path = H]{-5};   	 
	\addplot[domain=2002.25:2002.92, name path = I]{25};
	\addplot[domain=2002.25:2002.92, name path = J]{-5};   			
	\addplot[domain=2008.09:2010.5, name path = K]{25};
	\addplot[domain=2008.09:2010.5, name path = L]{-5};   		
	\addplot[lightgray] fill between[of=C and D];
	\addplot[lightgray] fill between[of=E and F];
	\addplot[lightgray] fill between[of=G and H];           
	\addplot[lightgray] fill between[of=I and J];
	\addplot[lightgray] fill between[of=K and L]; 		
	\nextgroupplot[title = Females]
	\addplot[black, very thick] table[x = year, y = extensive]{results_mean_female_2021_500.txt};
\addplot[black, dotted,  very thick] table[x = year, y = intensive]{results_mean_female_2021_500.txt};
\addplot[black, dashed, very thick] table[x = year, y = structural]{results_mean_female_2021_500.txt};		
\addplot[black, dashdotted, very thick] table[x = year, y = composition]{results_mean_female_2021_500.txt};
\addplot[black, double, very thick] table[x = year, y = total]{results_mean_female_2021_500.txt};				
	\addplot[domain=1981.0:1981.6, name path = C]{25};
	\addplot[domain=1981.0:1981.6, name path = D]{-5};      
	\addplot[domain=1982.6:1983.92, name path = E]{25};
	\addplot[domain=1982.6:1983.92, name path = F]{-5};      		
	\addplot[domain=1991.6:1992.25, name path = G]{25};
	\addplot[domain=1991.6:1992.25, name path = H]{-5};   	 
	\addplot[domain=2002.25:2002.92, name path = I]{25};
	\addplot[domain=2002.25:2002.92, name path = J]{-5};   			
	\addplot[domain=2008.09:2010.5, name path = K]{25};
	\addplot[domain=2008.09:2010.5, name path = L]{-5};   		
	\addplot[lightgray] fill between[of=C and D];
	\addplot[lightgray] fill between[of=E and F];
	\addplot[lightgray] fill between[of=G and H];           
	\addplot[lightgray] fill between[of=I and J];
	\addplot[lightgray] fill between[of=K and L]; 
	\end{groupplot}
	\node at (plots c2r1.east) [inner sep=10pt,anchor=north, xshift = 2cm, yshift=3cm] {\ref{grouplegend1}};  
	\end{tikzpicture}
\caption{Decomposition of the changes in mean annual earnings relative to
1976 with censoring at 500 hours.}
\label{fig:decomp_mean1}
\end{figure}

To explore the empirical impact of these lower hour individuals we redefined
individuals with under 500 annual hours as non participants.\label%
{page:robustness} We re-estimate our model and reproduce the decompositions
for the mean earnings. Because lower-hours individuals are not classified as
workers, the entering marginal workers enter at a higher level of hours (500
instead of 1). Figure \ref{fig:decomp_mean1} shows that the results for
males are almost identical to those for the zero hour cutoff.\footnote{%
This is explained by Figure \ref{fig:employment} of Fern\'{a}ndez.Val et al.
(2021b) which indicates that the majority of males are working a substantial
numbers of annual hours.} Thus, the choice of cut-off hours is less
influential in reassigning the impact across the two hours effects as the
\textquotedblleft entrants" are likely to be assigned a relatively high
number of annual hours and, more importantly, a number similar to those
already in employment. The corresponding figure for females reveals a
different story. For the zero cut-off approximately a quarter of the 80\%
increase in earnings over the sample period was due to the hours effects and
by 2019 approximately three quarters of this was due to movements on the
intensive margin and one quarter to the extensive margin. As the minimum of
an entrant is 500 hours the extensive margin effect increases and is
generally equal to that of the intensive margin.\footnote{%
This is explained by Figure \ref{fig:employment} which shows that a large
number of females work less than 500 hours.} In fact, by the end of the
sample period it is the most important of the two. Clearly, an exercise with
a higher cut-off point would attribute an even higher share to the extensive
margin. The sensitivity to the \textquotedblleft cut-off" point highlights
the importance of the definition of the extensive margin and intensive
margin effects. We prefer to conduct our analysis over the entire working
sample as the lower hours workers need to be included to appropriately
describe earnings inequality. They also provide a clearer understanding of
the role of hours on earnings inequality.

\pgfplotstableread{results_quantile_male_1_90_2020.txt}{\resultsa} %
\pgfplotstableread{results_quantile_male_2_90_2020.txt}{\resultsaa} %
\pgfplotstableread{results_quantile_male_3_90_2020.txt}{\resultsab} %
\pgfplotstableread{results_quantile_male_4_90_2020.txt}{\resultsac} %
\pgfplotstableread{results_quantile_male_5_90_2020.txt}{\resultsad}

\pgfplotstableread{results_quantile_female_1_90_2020.txt}{\resultsb} %
\pgfplotstableread{results_quantile_female_2_90_2020.txt}{\resultsba} %
\pgfplotstableread{results_quantile_female_3_90_2020.txt}{\resultsbb} %
\pgfplotstableread{results_quantile_female_4_90_2020.txt}{\resultsbc} %
\pgfplotstableread{results_quantile_female_5_90_2020.txt}{\resultsbd}


We employ two additional approaches to further explore this issue. First we
re-estimated the model and conducted the decompositions using the CPS Merged
Outgoing Rotation Group (MORG) data. The dependent variable is now weekly
earnings and the censoring variable is hours worked in the survey week. The
use of this measure of hours avoids the issue raised in our discussion of
the ASEC data which may inappropriately treat individuals working the same
annual hours, but not the same weekly hours, as identical. These data assign
a higher weight to the extensive margin as an individual not working in the
survey week will be recorded as non employed. For the ASEC data an
individual is recorded as employed even if they work only one week of the
year. The decomposition for male mean earnings is shown in Figure \ref%
{fig:decomp_morg} of Fern\'{a}ndez-Val et al. (2021b). Over the entire
period there is a reduction in weekly earnings of approximately 8\% which is
similar to the reduction in annual earnings. As expected the role of the
extensive margin is more important. For example, the biggest reduction in
weekly earnings occurs over the Great Recession. During this period there is
a large increase in contribution of the extensive margin. However the impact
of the two hours effects are approximately equal. Also the magnitude is
larger and for the weekly data there appears to be more drastic movement in
the extensive effect whereas for the annual data the response is flatter.
However, more interesting is that the dramatic effect in the intensive
margin effect exists for both data sets. The hours effects for female weekly
earnings shown in Figure \ref{fig:decomp_morg} of Fern\'andez-Val et al.
(2021b) are very similar to those for annual earnings. The trends and the
magnitudes are similar for both data sets and the primary difference is that
the annual data attributes more of the total hours effects to the intensive
margin.

\pgfplotstableread{results_mean_male_1_2020_alt.txt}{\resultsa} %
\pgfplotstableread{results_mean_male_2_2020_alt.txt}{\resultsaa} %
\pgfplotstableread{results_mean_male_3_2020_alt.txt}{\resultsab} %
\pgfplotstableread{results_mean_male_4_2020_alt.txt}{\resultsac} %
\pgfplotstableread{results_mean_male_5_2020_alt.txt}{\resultsad}

\pgfplotstableread{results_mean_female_1_2020_alt.txt}{\resultsb} %
\pgfplotstableread{results_mean_female_2_2020_alt.txt}{\resultsba} %
\pgfplotstableread{results_mean_female_3_2020_alt.txt}{\resultsbb} %
\pgfplotstableread{results_mean_female_4_2020_alt.txt}{\resultsbc} %
\pgfplotstableread{results_mean_female_5_2020_alt.txt}{\resultsbd}

\begin{figure}[tbp]
\centering
\begin{tikzpicture}
	\begin{groupplot}
	[
	group style={%
		group size = 2 by 1,
		group name=plots
		, horizontal sep=1.25cm, 
		xlabels at=edge bottom,
		ylabels at=edge left,
		x descriptions at=edge bottom
	},
	set layers,cell picture=true,
	width=0.38\textwidth,
	height=0.38\textwidth,
	legend columns=-1,
	xlabel = Survey year,
	ylabel = Relative difference,
	ymin = -0.4,
	ymax = 1.3,
	xmin = 1976, 
	xmax = 2019,
	cycle list name=black white
	]
	\nextgroupplot[legend to name=grouplegend1, title = Males, ymin = -0.2, ymax = 0.2]
	\addplot[black, very thick] table{\resultsa};
	\addlegendentry{Extensive}
	\addplot[black, dotted,  very thick] table{\resultsaa};
	\addlegendentry{Intensive}
	\addplot[black, dashed, very thick] table{\resultsab};
	\addlegendentry{Structural}	
	\addplot[black, dashdotted, very thick] table{\resultsac};
	\addlegendentry{Composition}		
	\addplot[black, double, very thick] table{\resultsad};		
	\addlegendentry{Total}		
	\addplot[domain=1981.0:1981.6, name path = C]{25};
	\addplot[domain=1981.0:1981.6, name path = D]{-5};      
	\addplot[domain=1982.6:1983.92, name path = E]{25};
	\addplot[domain=1982.6:1983.92, name path = F]{-5};      		
	\addplot[domain=1991.6:1992.25, name path = G]{25};
	\addplot[domain=1991.6:1992.25, name path = H]{-5};   	 
	\addplot[domain=2002.25:2002.92, name path = I]{25};
	\addplot[domain=2002.25:2002.92, name path = J]{-5};   			
	\addplot[domain=2008.09:2010.5, name path = K]{25};
	\addplot[domain=2008.09:2010.5, name path = L]{-5};   		
	\addplot[lightgray] fill between[of=C and D];
	\addplot[lightgray] fill between[of=E and F];
	\addplot[lightgray] fill between[of=G and H];           
	\addplot[lightgray] fill between[of=I and J];
	\addplot[lightgray] fill between[of=K and L]; 		
	\nextgroupplot[title = Females]
	\addplot[black, very thick] table{\resultsb};
	\addplot[black, dotted,  very thick] table{\resultsba};
	\addplot[black, dashed, very thick] table{\resultsbb};
	\addplot[black, dashdotted, very thick] table{\resultsbc};
	\addplot[black, double, very thick] table{\resultsbd};			
	\addplot[domain=1981.0:1981.6, name path = C]{25};
	\addplot[domain=1981.0:1981.6, name path = D]{-5};      
	\addplot[domain=1982.6:1983.92, name path = E]{25};
	\addplot[domain=1982.6:1983.92, name path = F]{-5};      		
	\addplot[domain=1991.6:1992.25, name path = G]{25};
	\addplot[domain=1991.6:1992.25, name path = H]{-5};   	 
	\addplot[domain=2002.25:2002.92, name path = I]{25};
	\addplot[domain=2002.25:2002.92, name path = J]{-5};   			
	\addplot[domain=2008.09:2010.5, name path = K]{25};
	\addplot[domain=2008.09:2010.5, name path = L]{-5};   		
	\addplot[lightgray] fill between[of=C and D];
	\addplot[lightgray] fill between[of=E and F];
	\addplot[lightgray] fill between[of=G and H];           
	\addplot[lightgray] fill between[of=I and J];
	\addplot[lightgray] fill between[of=K and L]; 
	\end{groupplot}
	\node at (plots c1r1.south) [inner sep=10pt,anchor=north, xshift= 3.5cm,yshift=-5ex] {\ref{grouplegend1}};  
	\end{tikzpicture}
\caption{Decomposition of the relative changes in mean annual earnings with
respect to 1976 - alternative method}
\label{fig:decomp_mean_alt}
\end{figure}
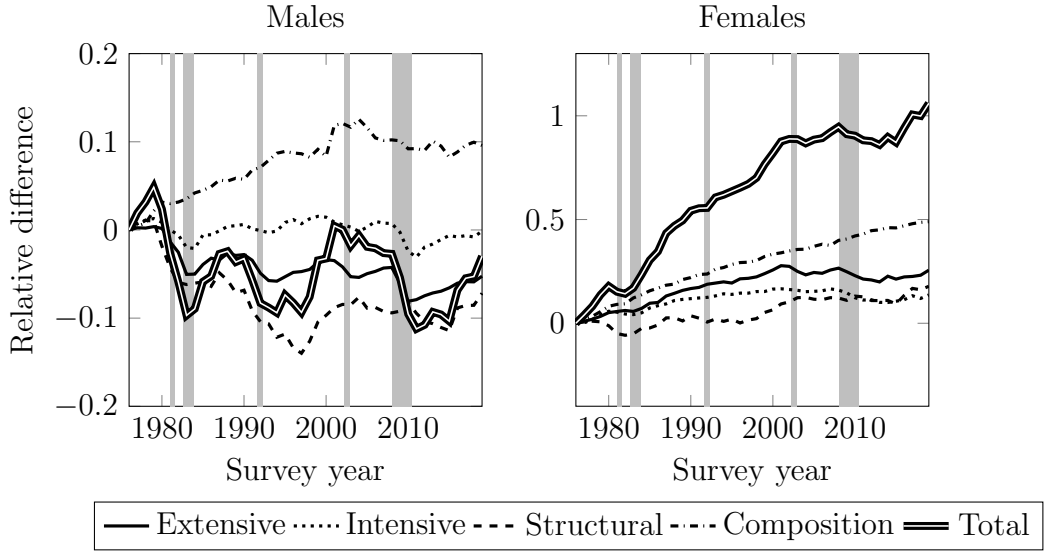

Another approach to evaluate the impact of non employment during the year on
the attribution to the two hours effects is to treat an observation working
part of the year as two different observations with weights reflecting the
fraction of the year worked.\footnote{%
We are grateful to a referee for this suggestion.} This also assigns a
greater weight to the extensive margin as any individual not employed 52
weeks per year is contributing to the extensive margin. This definition is
similar to that employed in Blundell et al. (2013).\label{insert:blundell_1}
The associated decomposition is shown in Figure \ref{fig:decomp_mean_alt}
and while the differences with the earlier figures can be anticipated the
relative magnitude of the hours effects are interesting. For males the
intensive margin is essentially zero for the whole sample except for a
notable negative impact during the Great Recession. In contrast the
extensive margin is larger and more closely tracks the overall decrease in
annual earnings. The impact at the extensive margin is much larger for the
Great Recession than the earlier figures although there remains a
substantial intensive margin effect for this period. For females the
extensive margin effect is now larger than the intensive margin effect for
the entire sample period although by 2019 the effects are close in
magnitude. This figure supports our earlier conjectures that while some of
the extensive margin is being allocated to the intensive margin there
appears a contribution from each.

These three alternative approaches highlight the nature of the sensitivity
of the relative magnitude of the extensive and intensive hours effects to
the modelling approach, and particularly the monotonicity assumption, and
the time unit associated with the measurement of employment. We acknowledge
that our approach is likely to underestimate the extensive margin effects
but substantial evidence of intensive margin effects remain even if we
reassign some part of them to the extensive margin.

\pgfplotstableread{results_quantile_male_1_mean_1.txt}{\resultsa} %
\pgfplotstableread{results_quantile_male_2_mean_1.txt}{\resultsaa} %
\pgfplotstableread{results_quantile_male_3_mean_1.txt}{\resultsab} %
\pgfplotstableread{results_quantile_male_4_mean_1.txt}{\resultsac} %
\pgfplotstableread{results_quantile_male_5_mean_1.txt}{\resultsad} %
\pgfplotstableread{results_quantile_male_6_mean_1.txt}{\resultsae} 

\pgfplotstableread{results_mean_male_1_2020_alternative1.txt}{\resultsa} %
\pgfplotstableread{results_mean_male_2_2020_alternative1.txt}{\resultsaa} %
\pgfplotstableread{results_mean_male_3_2020_alternative1.txt}{\resultsab} %
\pgfplotstableread{results_mean_male_4_2020_alternative1.txt}{\resultsac} %
\pgfplotstableread{results_mean_male_5_2020_alternative1.txt}{\resultsad}

\pgfplotstableread{results_mean_female_1_2020_alternative1.txt}{\resultsb} %
\pgfplotstableread{results_mean_female_2_2020_alternative1.txt}{\resultsba} %
\pgfplotstableread{results_mean_female_3_2020_alternative1.txt}{\resultsbb} %
\pgfplotstableread{results_mean_female_4_2020_alternative1.txt}{\resultsbc} %
\pgfplotstableread{results_mean_female_5_2020_alternative1.txt}{\resultsbd}

\subsection{Decomposing changes in earnings inequality}

\label{ss:inequality}

Studies of inequality typically examine the ratio of quantiles above and
below the median. Given the inclusion of those working zero hours in our
sample, the female participation rate of 56.5\% in 1976 does not allow
comparisons for females involving earnings below the fourth decile.
Accordingly, we begin in Figure~\ref{fig:d5d9} by focusing on relative
changes in the D9/D5 ratio.

\pgfplotstableread{results_quantile_male_1_d9d5.txt}{\resultsaa} %
\pgfplotstableread{results_quantile_male_2_d9d5.txt}{\resultsab} %
\pgfplotstableread{results_quantile_male_3_d9d5.txt}{\resultsac} %
\pgfplotstableread{results_quantile_male_4_d9d5.txt}{\resultsad} %
\pgfplotstableread{results_quantile_male_5_d9d5.txt}{\resultsae}

\pgfplotstableread{results_quantile_female_1_d9d5.txt}{\resultsa} %
\pgfplotstableread{results_quantile_female_2_d9d5.txt}{\resultsb} %
\pgfplotstableread{results_quantile_female_3_d9d5.txt}{\resultsc} %
\pgfplotstableread{results_quantile_female_4_d9d5.txt}{\resultsd} %
\pgfplotstableread{results_quantile_female_5_d9d5.txt}{\resultse}

\begin{figure}[tbp]
\centering
\begin{tikzpicture}
	\begin{groupplot}
	[
	group style={%
		group size = 2 by 1,
		group name=plots
		, horizontal sep=1.5cm, 
		xlabels at=edge bottom,
		ylabels at=edge left,
		x descriptions at=edge bottom
	},
	set layers,cell picture=true,
	width=0.4\textwidth,
	height=0.4\textwidth,
	legend columns=1,
	xlabel = Survey year,
	ylabel = Relative difference,
	ymin = -0.8,
	ymax = 0.7,
	xmin = 1976, 
	xmax = 2019,
	cycle list name=black white
	]
	\nextgroupplot[legend to name=grouplegend1,title=Males]
	\addplot[black, very thick] table {\resultsaa};
	\addlegendentry{Extensive}
	\addplot[black, very thick, dotted] table {\resultsab};
	\addlegendentry{Intensive}
	\addplot[black, very thick, dashed] table {\resultsac};
	\addlegendentry{Structural}
	\addplot[black, very thick, dashdotted] table {\resultsad};
	\addlegendentry{Composition}
	\addplot[black, very thick, double] table {\resultsae};	
	\addlegendentry{Total}			
         \addplot[domain=1981.0:1981.6, name path = C]{25};
         \addplot[domain=1981.0:1981.6, name path = D]{-5};      
         \addplot[domain=1982.6:1983.92, name path = E]{25};
	\addplot[domain=1982.6:1983.92, name path = F]{-5};      		
         \addplot[domain=1991.6:1992.25, name path = G]{25};
         \addplot[domain=1991.6:1992.25, name path = H]{-5};   	 
         \addplot[domain=2002.25:2002.92, name path = I]{25};
         \addplot[domain=2002.25:2002.92, name path = J]{-5};   			
         \addplot[domain=2008.09:2010.5, name path = K]{25};
	\addplot[domain=2008.09:2010.5, name path = L]{-5};   		
         \addplot[lightgray] fill between[of=C and D];
         \addplot[lightgray] fill between[of=E and F];
         \addplot[lightgray] fill between[of=G and H];           
         \addplot[lightgray] fill between[of=I and J];
         \addplot[lightgray] fill between[of=K and L]; 
	\nextgroupplot[title=Females]
	\addplot[black, very thick] table {\resultsa};
	\addplot[black, very thick, dotted] table {\resultsb};
	\addplot[black, very thick, dashed] table {\resultsc};
	\addplot[black, very thick, dashdotted] table {\resultsd};
	\addplot[black, very thick, double] table {\resultse};
         \addplot[domain=1981.0:1981.6, name path = C]{25};
         \addplot[domain=1981.0:1981.6, name path = D]{-5};      
         \addplot[domain=1982.6:1983.92, name path = E]{25};
	\addplot[domain=1982.6:1983.92, name path = F]{-5};      		
         \addplot[domain=1991.6:1992.25, name path = G]{25};
         \addplot[domain=1991.6:1992.25, name path = H]{-5};   	 
         \addplot[domain=2002.25:2002.92, name path = I]{25};
         \addplot[domain=2002.25:2002.92, name path = J]{-5};   			
         \addplot[domain=2008.09:2010.5, name path = K]{25};
	\addplot[domain=2008.09:2010.5, name path = L]{-5};   		
         \addplot[lightgray] fill between[of=C and D];
         \addplot[lightgray] fill between[of=E and F];
         \addplot[lightgray] fill between[of=G and H];           
         \addplot[lightgray] fill between[of=I and J];
         \addplot[lightgray] fill between[of=K and L]; 
	\end{groupplot}
\node at (plots c2r1.east) [inner sep=10pt,anchor=north, xshift = 2cm, yshift=3cm] {\ref{grouplegend1}}; 
	\end{tikzpicture}
\caption{Decomposition of the changes relative to 1976 in the ratio of the
upper decile to the median of annual earnings.}
\label{fig:d5d9}
\end{figure}
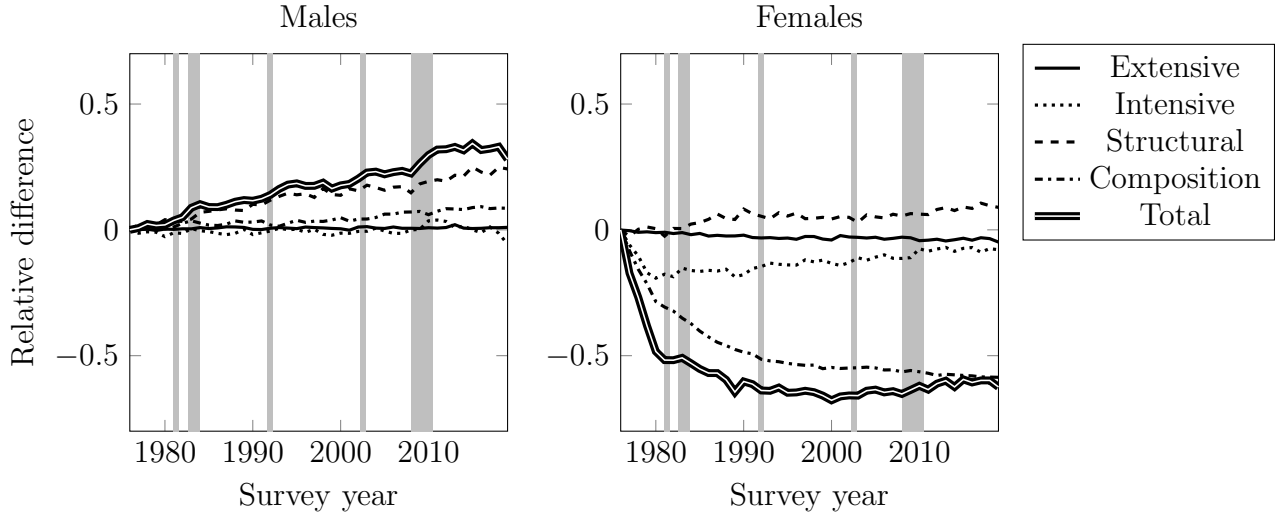


The 28\% increase for males is substantial and economically significant.
There are no hours effects and the increase largely reflects the structural
wage component. Existing empirical work has found substantial and positive
real wage growth at the higher quantiles while real wages below Q3 have
decreased. This reflects the increase in the returns to education at the
higher quantiles of the wage distribution. There is evidence of small
composition effects as the effects operating at the two quantiles offset
each other. The hours effects are small and consistent with earlier research
which finds working hours have a relatively modest impact in explaining the
changes in males earnings inequality in the United States (see, for example,
Blau and Kahn, 2009 and Checchi et al, 2016). Movements on the extensive
margin at both the median and the ninth decile are negligible and have no
effect on this measure of inequality.

The corresponding figure for females differs in several ways. First, there
is a drastic decrease of over 60\% in this measure of inequality. The fall
occurs primarily from 1976 to 1990 and is maintained for the remainder of
the sample period. This large effect is initially due to the composition and
intensive hours effects. For the first part of the sample period both
effects are contributing to the large decrease in inequality. However, by
the mid 1980s the intensive margin effect begins to diminish and the large
decrease in inequality observed at the end of the sample period is almost
entirely due to the composition effect. The intensive margin appears to
capture the increased hours of those working relatively few hours early in
the sample period who subsequently increase their hours in latter years. The
presence of an hours effect differs from the conclusions of Blau and Kahn
(2011) and Checchi (2016) who report that the variance in working hours is
relatively small in the U.S. and the change in the correlation between
working hours and wages is small. This reflects the benefit of our approach
which isolates movements in hours at specific points of the earnings
distribution rather than employing the variance based on the whole
distribution. Also note that these studies focus only on those working
positive hours. Second, the structural wage effect is relatively unimportant
for this measure of female earnings inequality. The evidence in FPVV
suggests the structural wage effects should be increasing inequality as
those at the upper quantiles of the wage distribution have experienced much
larger wage increases from this source than those lower in the wage
distribution. There is some evidence in support of this phenomena. The
extensive hours effect appears to be non existent. \label%
{page:comment_median}

We conclude our empirical work by examining the interquartile range for
males shown in Figure \ref{fig:decomp_q3q1}. While the magnitude of its
increase is not greatly different to the earlier measure its evolution over
the time period is drastically different. While the D9/D5 ratio shows a
steady increase, the interquartile ratio increases in large jumps during the
recessionary period. The Great Recession generates an increase in the Q3/Q1
from 20\% to 60\%. This has been followed by an equally dramatic decrease in
inequality for the remainder of our sample period although it remains well
above that before the Great Recession. The trend in the inequality can be
almost fully described by the structural effect although it is partially
explained by the composition effect since the 2000s. This is consistent with
the earlier figures which illustrate the negative impact of the composition
effect on earnings at Q1 for the same period. The cyclicality in the Q3/Q1
ratio is also partially explained by the two hours effects. This is most
apparent in the Great Recession which sees a large increase in the intensive
hours effect.%



\pgfplotstableread{results_quantile_male_1_q3q1.txt}{\resultsa} %
\pgfplotstableread{results_quantile_male_2_q3q1.txt}{\resultsaa} %
\pgfplotstableread{results_quantile_male_3_q3q1.txt}{\resultsab} %
\pgfplotstableread{results_quantile_male_4_q3q1.txt}{\resultsac} %
\pgfplotstableread{results_quantile_male_5_q3q1.txt}{\resultsad} %
\pgfplotstableread{results_quantile_male_6_q3q1.txt}{\resultsae}

\begin{figure}[tbp]
\centering
\begin{tikzpicture}
	\begin{groupplot}
	[
	group style={%
		group size = 2 by 1,
		group name=plots
		, horizontal sep=1.25cm, 
		xlabels at=edge bottom,
		ylabels at=edge left,
		x descriptions at=edge bottom
	},
	set layers,cell picture=true,
	width=0.4\textwidth,
	height=0.4\textwidth,
	legend columns=1,
	xlabel = Survey year,
	ylabel = Relative difference,
	ymin = -0.2,
	ymax = 0.7,
	xmin = 1976, 
	xmax = 2019,
	cycle list name=black white
	]
	\nextgroupplot[legend to name=grouplegend1, title = Males]
	\addplot[black, very thick] table{\resultsa};
	\addlegendentry{Extensive}
	\addplot[black, dotted,  very thick] table{\resultsaa};
	\addlegendentry{Intensive}
	\addplot[black, dashed, very thick] table{\resultsab};
	\addlegendentry{Structural}	
	\addplot[black, dashdotted, very thick] table{\resultsac};
	\addlegendentry{Composition}		
	\addplot[black, double, very thick] table{\resultsad};		
	\addlegendentry{Total}		
	\addplot[domain=1981.0:1981.6, name path = C]{25};
	\addplot[domain=1981.0:1981.6, name path = D]{-5};      
	\addplot[domain=1982.6:1983.92, name path = E]{25};
	\addplot[domain=1982.6:1983.92, name path = F]{-5};      		
	\addplot[domain=1991.6:1992.25, name path = G]{25};
	\addplot[domain=1991.6:1992.25, name path = H]{-5};   	 
	\addplot[domain=2002.25:2002.92, name path = I]{25};
	\addplot[domain=2002.25:2002.92, name path = J]{-5};   			
	\addplot[domain=2008.09:2010.5, name path = K]{25};
	\addplot[domain=2008.09:2010.5, name path = L]{-5};   		
	\addplot[lightgray] fill between[of=C and D];
	\addplot[lightgray] fill between[of=E and F];
	\addplot[lightgray] fill between[of=G and H];           
	\addplot[lightgray] fill between[of=I and J];
	\addplot[lightgray] fill between[of=K and L]; 		
	\end{groupplot}
	\node at (plots c1r1.east) [inner sep=10pt,anchor=north, xshift= 3.5cm, yshift=3cm] {\ref{grouplegend1}};  
	\end{tikzpicture}
\caption{Decomposition of the changes in the Q3/Q1 ratio relative to 1976.}
\label{fig:decomp_q3q1}
\end{figure}
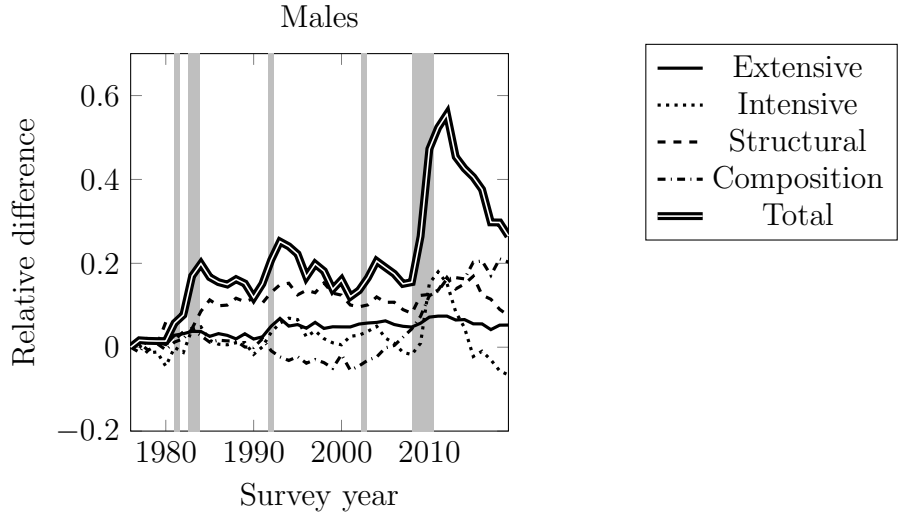

The inequality decompositions illustrate the benefit of examining earnings
inequality and specifically the impact operating through hours.\label%
{insert:economics_1} In contrast to wage inequality, the results differ
greatly by gender. The most striking finding for males is the remarkable
sensitivity of inequality to recessionary factors. While there is evidence
of this also in wages it is far more dramatic in earnings. The total hours
effects in the inequality decomposition are greater than the structural
effects responsible for wage inequality. The increases in earnings
inequality are primarily generated by reduction in hours worked by the lower
wage males during recessions. While it is beyond the scope of this paper to
provide policy prescriptions to address earnings inequality the implications
of our results are clear. Lower wage males suffer greatly in recessionary
periods through both lower wages and lower demand for their labor. The lower
level of demand is reflected in spells of unemployment and reduction in
working hours for those employed. This suggests that the appropriate
policies for combatting earnings inequality should focus on supplementing
the lower wage males in these periods. This could take the form of targeted
unemployment programs or through incentives for employers to maintain the
hours of these workers through recessions. The results for females are
remarkably different. Although female wage inequality has increased in a
manner similarly to males, female earnings inequality has decreased
dramatically. This has occurred despite changes in hours affecting both the
earnings of low and high earning females. However, the larger decreases at
the bottom have reduced earnings inequality. This also has very clear policy
implications for reducing inequality. Policies should focus on enabling
females to increase their level of labor supply.

While our reduced form approach highlights the relationship between hours
and earnings, we acknowledge that effective public policy is guided by a
deeper understanding of individuals' economic motivations. While we
establish that movements on both the extensive and intensive margins are
important, understanding why they occur requires an approach imposing
greater economic structure. For example, Keane and Wasi (2016) estimate
supply elasticities which incorporate both margins and investigate how they
vary by individual characteristics such as age and education. Policies based
on our findings in conjunction with those from studies such as Keane and
Wasi (2016) are more likely to achieve goals such as reducing earnings
inequality.\label{insert:keane}

\section{Conclusions}

\label{sec:conclusions}

This paper examines the relationship between annual hours of work and annual
earnings by decomposing movements in the annual earnings distribution into
composition, structural and hours effects. This requires the estimation of a
nonseparable model of hours and wages in the presence of a censored
selection rule. Our empirical work has a number of interesting results.
Changes in the hours of annual work are important for the lower quartile of
male annual earnings but their importance quickly diminishes as we examine
higher quantiles. Using the interquartile ratio as an alternative measure of
inequality reveals that hours effects are very important and particularly
notable in recessionary periods. Changes in the female distribution of
annual hours have increased earnings at all quantiles of the female annual
earnings. These effects are generally operating through the intensive margin
although there is also an important role for the extensive margin. The
female upper decile to median ratio decreases dramatically indicating a
drastic reduction in earnings inequality. This reduction is due to large
effects at the intensive margin of annual hours of work but also composition
effects, such as increased levels of education, which increase wages at all
quantiles and increase the level of annual hours of work.

\newpage

\begin{center}
\textbf{\Large{Hours worked and the U.S. distribution of real annual earnings
1976--2019, Web Appendix}}\\

\medskip

Ivan Fern\'andez-Val, Aico van Vuuren, Francis Vella, \\ and Franco Peracchi

\end{center}

\appendix
\section{Proof of proposition \protect\ref{prop:id}}

\label{web:id}
\begin{proof}[Proof of Proposition \protect\ref{prop:id} ]
{\small We focus on the identification of $G_{\langle q,r,s,t\rangle }$. The
identification of $G_{\langle r,s,t\rangle }$ is a special case. }

{\small The rank invariance $V^1_q = V^2_r$ is needed to replace the
distribution $F_{X_{p},Z_{p},V_{p}^{1},V_{p}^{2}}$ by $F_{X_{p},Z_{p},V_{p}}$
in the integrals in \eqref{eq:g_qrst2}. Both integrals are then zero outside
the support of $F_{X_{p},Z_{p},V_{p}}$. The areas of integration are
expressed in terms of the function $k_q$. These areas are well-defined
provided that $\mathcal{XZV}_{p}\subseteq \mathcal{XZV}_{q}$. This condition
simplifies to $\mathcal{XZ}_{p}\subseteq \mathcal{XZ}_{q}$ because $%
k_{q}(x,z,v)=Q_{H_{q}\mid X_{q},Z_{q}}(v\mid x,z):=\inf \{h\in \mathbb{R}%
:F_{H_{q}\mid X_{q},Z_{q}}(h\mid x,z)\geq v\}$ is monotone nondecreasing in $%
v$, so that we can set $k^{q}(x,z,v)=0$ if $v \in \mathcal{V}_p$ but $v
\notin \mathcal{V}_q$. }

{\small Next, we analyze the integrand of the first integral. The LDSF $%
G_{s}(\cdot ,x,v)$ is identified over the support $\mathcal{XV}_{s}^{\ast }$%
, while the area of integration is $\mathcal{XZV}_{p}\cap \mathcal{XZV}%
_{q}^{\ast }$. This yields the condition $\mathcal{XV}_{p}\cap \mathcal{XV}%
_{q}^{\ast }\subseteq \mathcal{XV}_{s}^{\ast }$. The function $k_{r}(x,z,v)$
appears in the first argument of $G_{s}(w,x,v)$. This function $k_{r}(x,z,v)$
is identified for $(x,z)\in \mathcal{XZ}_{r}$ and any $v$ by monotonicity,%
\footnote{%
As above, we can set $k^{r}(x,z,v)=0$ if $v\notin \mathcal{V}_{r}$.} which
combined with the area of integration yields the condition $\mathcal{XZ}%
_{q}^{\ast }\cap \mathcal{XZ}_{p}\subseteq \mathcal{XZ}_{r}$. 
}
\end{proof}

\section{Figures}

\label{web:figures}

\counterwithin{figure}{section} \setcounter{figure}{0} 
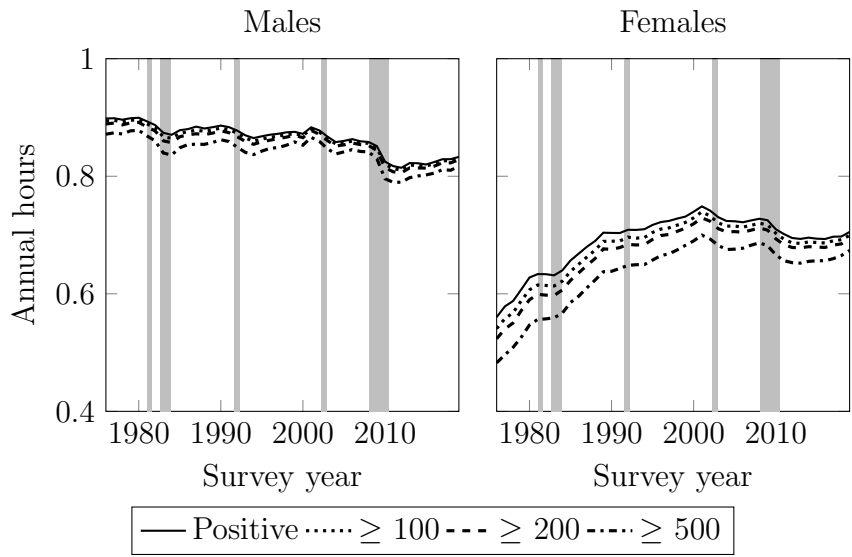
\begin{figure}[h!]
\centering
\begin{tikzpicture}
	\begin{groupplot}
	[
	group style={%
		vertical sep = 3cm,
		group size = 2 by 1,
		group name=plots,
		horizontal sep=0.5cm, 
		xlabels at=edge bottom,
		y descriptions at=edge left,
		ylabels at=edge left,
		x descriptions at=edge bottom
	},
	set layers,cell picture=true,
	width=0.38\textwidth,
	height=0.38\textwidth,
	legend columns=-1,
	xlabel = Survey year,
	ylabel = Annual hours,
	ymin = 0.4,
	ymax = 1,
	xmin = 1976, 
	xmax = 2019,
	cycle list name=black white
	]	
	\nextgroupplot[legend to name=grouplegend1,title = Males]
	\addplot[black, thick] table[x = year, y = zero]{employment_males1.txt};
	\addlegendentry{Positive}
	\addplot[black, very thick, dotted] table[x = year, y = hours100]{employment_males1.txt};
	\addlegendentry{$\geq$ 100}
	\addplot[black, very thick, dashed] table[x = year, y = hours200]{employment_males1.txt};
	\addlegendentry{$\geq$ 200}	
	\addplot[black, very thick, dashdotted] table[x = year, y = hours500]{employment_males1.txt};	
	\addlegendentry{$\geq$ 500}	
	\addplot[domain=1981.0:1981.6, name path = C]{1};
	\addplot[domain=1981.0:1981.6, name path = D]{0};      
	\addplot[domain=1982.6:1983.92, name path = E]{1};
	\addplot[domain=1982.6:1983.92, name path = F]{0};      		
	\addplot[domain=1991.6:1992.25, name path = G]{1};
	\addplot[domain=1991.6:1992.25, name path = H]{0};   	 
	\addplot[domain=2002.25:2002.92, name path = I]{1};
	\addplot[domain=2002.25:2002.92, name path = J]{0};   			
	\addplot[domain=2008.09:2010.5, name path = K]{1};
	\addplot[domain=2008.09:2010.5, name path = L]{0};   		
	\addplot[lightgray] fill between[of=C and D];
	\addplot[lightgray] fill between[of=E and F];
	\addplot[lightgray] fill between[of=G and H];           
	\addplot[lightgray] fill between[of=I and J];
	\addplot[lightgray] fill between[of=K and L]; 		
	\nextgroupplot[title = Females]
	\addplot[black, thick] table[x = year, y = zero]{employment_females1.txt};
	\addplot[black, very thick, dotted] table[x = year, y = hours100]{employment_females1.txt};
	\addplot[black, very thick, dashed] table[x = year, y = hours200]{employment_females1.txt};
	\addplot[black, very thick, dashdotted] table[x = year, y = hours500]{employment_females1.txt};	
	\addplot[domain=1981.0:1981.6, name path = C]{1};
	\addplot[domain=1981.0:1981.6, name path = D]{0};      
	\addplot[domain=1982.6:1983.92, name path = E]{1};
	\addplot[domain=1982.6:1983.92, name path = F]{0};      		
	\addplot[domain=1991.6:1992.25, name path = G]{1};
	\addplot[domain=1991.6:1992.25, name path = H]{0};   	 
	\addplot[domain=2002.25:2002.92, name path = I]{1};
	\addplot[domain=2002.25:2002.92, name path = J]{0};   			
	\addplot[domain=2008.09:2010.5, name path = K]{1};
	\addplot[domain=2008.09:2010.5, name path = L]{0};   		
	\addplot[lightgray] fill between[of=C and D];
	\addplot[lightgray] fill between[of=E and F];
	\addplot[lightgray] fill between[of=G and H];           
	\addplot[lightgray] fill between[of=I and J];
	\addplot[lightgray] fill between[of=K and L]; 	
	\end{groupplot}
	\node at (plots c1r1.south) [inner sep=10pt,anchor=north, xshift= 2cm,yshift=-5ex] {\ref{grouplegend1}};  
	\end{tikzpicture}
\caption{Time profile of the percentage of individuals with at least 1,
100, 200 and 500 working hours per year.}
\label{fig:employment}
\end{figure}

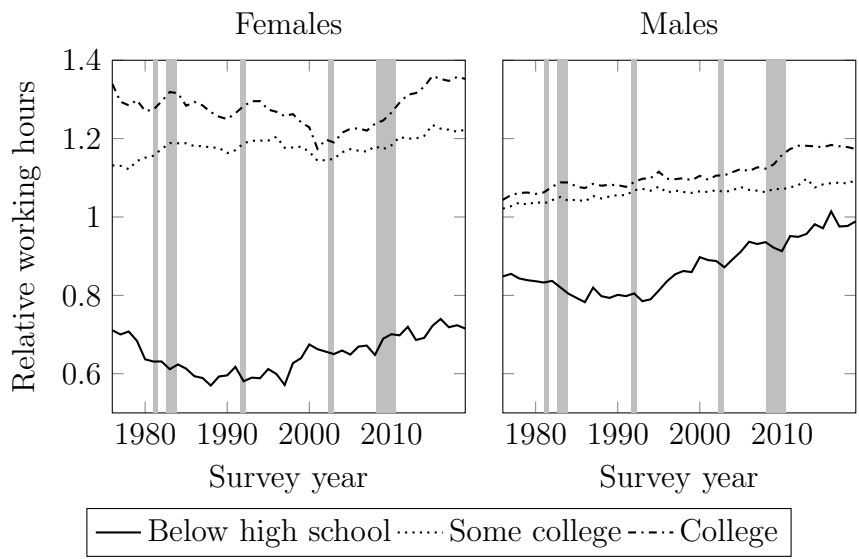
\begin{figure}[h!]
\centering
\begin{tikzpicture}
	\begin{groupplot}
	[
	group style={%
		group size = 2 by 1,
		group name=plots,
		horizontal sep=0.5cm, 
		xlabels at=edge bottom,
		y descriptions at=edge left,
		ylabels at=edge left,
		x descriptions at=edge bottom
	},
	set layers,cell picture=true,
	width=0.38\textwidth,
	height=0.38\textwidth,
	legend columns=-1,
	xlabel = Survey year,
	ylabel = Relative working hours,
	ymin = 0.5,
	ymax = 1.4,
	xmin = 1976, 
	xmax = 2019,
	cycle list name=black white
	]	
	\nextgroupplot[legend to name=grouplegend1, title = Females]			
	\addplot[black, thick] table[x = year, y = below]{relative_work_females_educ_2020.txt};
	\addlegendentry{Below high school}
	\addplot[black, thick, dotted] table[x = year, y = some]{relative_work_females_educ_2020.txt};
	\addlegendentry{Some college}	
	\addplot[black, thick, dashdotted] table[x = year, y = college]{relative_work_females_educ_2020.txt};
	\addlegendentry{College}	
	\addplot[domain = 1976:2016, black, dashed, thick]{0};
	\addplot[domain=1981.0:1981.6, name path = C]{2};
	\addplot[domain=1981.0:1981.6, name path = D]{0};      
	\addplot[domain=1982.6:1983.92, name path = E]{2};
	\addplot[domain=1982.6:1983.92, name path = F]{0};      		
	\addplot[domain=1991.6:1992.25, name path = G]{2};
	\addplot[domain=1991.6:1992.25, name path = H]{0};   	 
	\addplot[domain=2002.25:2002.92, name path = I]{2};
	\addplot[domain=2002.25:2002.92, name path = J]{0};   			
	\addplot[domain=2008.09:2010.5, name path = K]{2};
	\addplot[domain=2008.09:2010.5, name path = L]{0};   		
	\addplot[lightgray] fill between[of=C and D];
	\addplot[lightgray] fill between[of=E and F];
	\addplot[lightgray] fill between[of=G and H];           
	\addplot[lightgray] fill between[of=I and J];
	\addplot[lightgray] fill between[of=K and L]; 		
	\nextgroupplot[title = Males]	
	
	\addplot[black, thick] table[x = year, y = below]{relative_work_males_educ_2020.txt};
	\addplot[black, thick, dotted] table[x = year, y = some]{relative_work_males_educ_2020.txt};
	\addplot[black, thick, dashdotted] table[x = year, y = college]{relative_work_males_educ_2020.txt};
	\addplot[domain = 1976:2016, black, dashed, thick]{0};
	\addplot[domain=1981.0:1981.6, name path = C]{2};
	\addplot[domain=1981.0:1981.6, name path = D]{0};      
	\addplot[domain=1982.6:1983.92, name path = E]{2};
	\addplot[domain=1982.6:1983.92, name path = F]{0};      		
	\addplot[domain=1991.6:1992.25, name path = G]{2};
	\addplot[domain=1991.6:1992.25, name path = H]{0};   	 
	\addplot[domain=2002.25:2002.92, name path = I]{2};
	\addplot[domain=2002.25:2002.92, name path = J]{0};   			
	\addplot[domain=2008.09:2010.5, name path = K]{2};
	\addplot[domain=2008.09:2010.5, name path = L]{0};   		
	\addplot[lightgray] fill between[of=C and D];
	\addplot[lightgray] fill between[of=E and F];
	\addplot[lightgray] fill between[of=G and H];           
	\addplot[lightgray] fill between[of=I and J];
	\addplot[lightgray] fill between[of=K and L]; 		
	\end{groupplot}	
	\node at (plots c1r1.south) [inner sep=10pt,anchor=north, xshift= 2cm,yshift=-5ex] {\ref{grouplegend1}};  	
	\end{tikzpicture}
\caption{Employment levels for different education levels relative to high
school graduates.}
\label{fig:education_participation}
\end{figure}

\begin{figure}[h!]
\centering
\begin{tikzpicture}
	\begin{groupplot}
	[
	group style={%
		group size = 2 by 1,
		group name=plots,
		horizontal sep=0.5cm, 
		xlabels at=edge bottom,
		y descriptions at=edge left,
		ylabels at=edge left,
		x descriptions at=edge bottom
	},
	set layers,cell picture=true,
	width=0.38\textwidth,
	height=0.38\textwidth,
	legend columns=-1,
	xlabel = Survey year,
	ylabel = Relative working hours,
	ymin = 0.8,
	ymax = 1.2,
	xmin = 1976, 
	xmax = 2019,
	cycle list name=black white
	]	
		\nextgroupplot[title = Males]	
	\addplot[black, thick] table[x = year, y = below]{hours_relative_males_educ_2020.txt};
	\addplot[black, thick, dotted] table[x = year, y = some]{hours_relative_males_educ_2020.txt};
	\addplot[black, thick, dashdotted] table[x = year, y = college]{hours_relative_males_educ_2020.txt};
	\addplot[domain = 1976:2016, black, dashed, thick]{0};
	\addplot[domain=1981.0:1981.6, name path = C]{2};
	\addplot[domain=1981.0:1981.6, name path = D]{0};      
	\addplot[domain=1982.6:1983.92, name path = E]{2};
	\addplot[domain=1982.6:1983.92, name path = F]{0};      		
	\addplot[domain=1991.6:1992.25, name path = G]{2};
	\addplot[domain=1991.6:1992.25, name path = H]{0};   	 
	\addplot[domain=2002.25:2002.92, name path = I]{2};
	\addplot[domain=2002.25:2002.92, name path = J]{0};   			
	\addplot[domain=2008.09:2010.5, name path = K]{2};
	\addplot[domain=2008.09:2010.5, name path = L]{0};   		
	\addplot[lightgray] fill between[of=C and D];
	\addplot[lightgray] fill between[of=E and F];
	\addplot[lightgray] fill between[of=G and H];           
	\addplot[lightgray] fill between[of=I and J];
	\addplot[lightgray] fill between[of=K and L]; 		
	\nextgroupplot[legend to name=grouplegend1, title = Females]			
	\addplot[black, thick] table[x = year, y = below]{hours_relative_females_educ_2020.txt};
	\addlegendentry{Below high school}
	\addplot[black, thick, dotted] table[x = year, y = some]{hours_relative_females_educ_2020.txt};
	\addlegendentry{Some college}	
	\addplot[black, thick, dashdotted] table[x = year, y = college]{hours_relative_females_educ_2020.txt};
	\addlegendentry{College}	
	\addplot[domain = 1976:2016, black, dashed, thick]{0};
	\addplot[domain=1981.0:1981.6, name path = C]{2};
	\addplot[domain=1981.0:1981.6, name path = D]{0};      
	\addplot[domain=1982.6:1983.92, name path = E]{2};
	\addplot[domain=1982.6:1983.92, name path = F]{0};      		
	\addplot[domain=1991.6:1992.25, name path = G]{2};
	\addplot[domain=1991.6:1992.25, name path = H]{0};   	 
	\addplot[domain=2002.25:2002.92, name path = I]{2};
	\addplot[domain=2002.25:2002.92, name path = J]{0};   			
	\addplot[domain=2008.09:2010.5, name path = K]{2};
	\addplot[domain=2008.09:2010.5, name path = L]{0};   		
	\addplot[lightgray] fill between[of=C and D];
	\addplot[lightgray] fill between[of=E and F];
	\addplot[lightgray] fill between[of=G and H];           
	\addplot[lightgray] fill between[of=I and J];
	\addplot[lightgray] fill between[of=K and L]; 		
	\end{groupplot}	
	\node at (plots c1r1.south) [inner sep=10pt,anchor=north, xshift= 2cm,yshift=-5ex] {\ref{grouplegend1}};  	
	\end{tikzpicture}
\caption{Relative number of working hours for different education levels.
Working hours are relative to high school graduates.}
\label{fig:education_hours}
\end{figure}
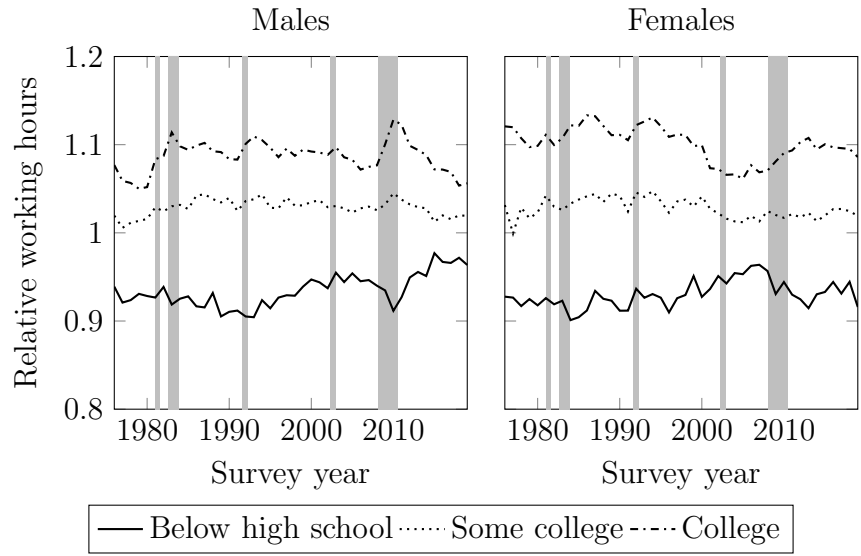

\begin{figure}[h!]
\centering
\begin{tikzpicture}
	\begin{groupplot}
	[
	group style={%
		group size = 2 by 1,
		group name=plots,
		horizontal sep=0.5cm, 
		xlabels at=edge bottom,
		y descriptions at=edge left,
		ylabels at=edge left,
		x descriptions at=edge bottom
	},
	set layers,cell picture=true,
	width=0.38\textwidth,
	height=0.38\textwidth,
	legend columns=-1,
	xlabel = Survey year,
	ylabel = Relative working hours,
	ymin = 0.7,
	ymax = 1.2,
	xmin = 1976, 
	xmax = 2019,
	cycle list name=black white
	]	
		\nextgroupplot[title = Males]	
	\addplot[black, thick] table[x = year, y = relative]{hours_relative_males_married_2020.txt};
	\addplot[black, thick, dotted] table[x = year, y = relative]{hours_relative_males_nchild_2020.txt};
	\addplot[black, thick, dashdotted] table[x = year, y = relative]{hours_relative_males_nchild5_2020.txt};
	\addplot[domain = 1976:2016, black, dashed, thick]{0};
	\addplot[domain=1981.0:1981.6, name path = C]{2};
	\addplot[domain=1981.0:1981.6, name path = D]{0};      
	\addplot[domain=1982.6:1983.92, name path = E]{2};
	\addplot[domain=1982.6:1983.92, name path = F]{0};      		
	\addplot[domain=1991.6:1992.25, name path = G]{2};
	\addplot[domain=1991.6:1992.25, name path = H]{0};   	 
	\addplot[domain=2002.25:2002.92, name path = I]{2};
	\addplot[domain=2002.25:2002.92, name path = J]{0};   			
	\addplot[domain=2008.09:2010.5, name path = K]{2};
	\addplot[domain=2008.09:2010.5, name path = L]{0};   		
	\addplot[lightgray] fill between[of=C and D];
	\addplot[lightgray] fill between[of=E and F];
	\addplot[lightgray] fill between[of=G and H];           
	\addplot[lightgray] fill between[of=I and J];
	\addplot[lightgray] fill between[of=K and L]; 		
	\nextgroupplot[legend to name=grouplegend1, title = Females]			
	\addplot[black, thick] table[x = year, y = relative]{hours_relative_females_married_2020.txt};
	\addlegendentry{Married}
	\addplot[black, thick, dotted] table[x = year, y = relative]{hours_relative_females_nchild_2020.txt};
	\addlegendentry{Child present}	
	\addplot[black, thick, dashdotted] table[x = year, y = relative]{hours_relative_females_nchild5_2020.txt};
	\addlegendentry{Child below 5 y/o age present}	
	\addplot[domain = 1976:2016, black, dashed, thick]{0};
	\addplot[domain=1981.0:1981.6, name path = C]{2};
	\addplot[domain=1981.0:1981.6, name path = D]{0};      
	\addplot[domain=1982.6:1983.92, name path = E]{2};
	\addplot[domain=1982.6:1983.92, name path = F]{0};      		
	\addplot[domain=1991.6:1992.25, name path = G]{2};
	\addplot[domain=1991.6:1992.25, name path = H]{0};   	 
	\addplot[domain=2002.25:2002.92, name path = I]{2};
	\addplot[domain=2002.25:2002.92, name path = J]{0};   			
	\addplot[domain=2008.09:2010.5, name path = K]{2};
	\addplot[domain=2008.09:2010.5, name path = L]{0};   		
	\addplot[lightgray] fill between[of=C and D];
	\addplot[lightgray] fill between[of=E and F];
	\addplot[lightgray] fill between[of=G and H];           
	\addplot[lightgray] fill between[of=I and J];
	\addplot[lightgray] fill between[of=K and L]; 		
	\end{groupplot}	
	\node at (plots c1r1.south) [inner sep=10pt,anchor=north, xshift= 2cm,yshift=-5ex] {\ref{grouplegend1}};  	
	\end{tikzpicture}
\caption{Relative number of working hours. Working hours are relative to
either unmarried, no children or no children under the age of 5.}
\label{fig:relative_hours_married}
\end{figure}
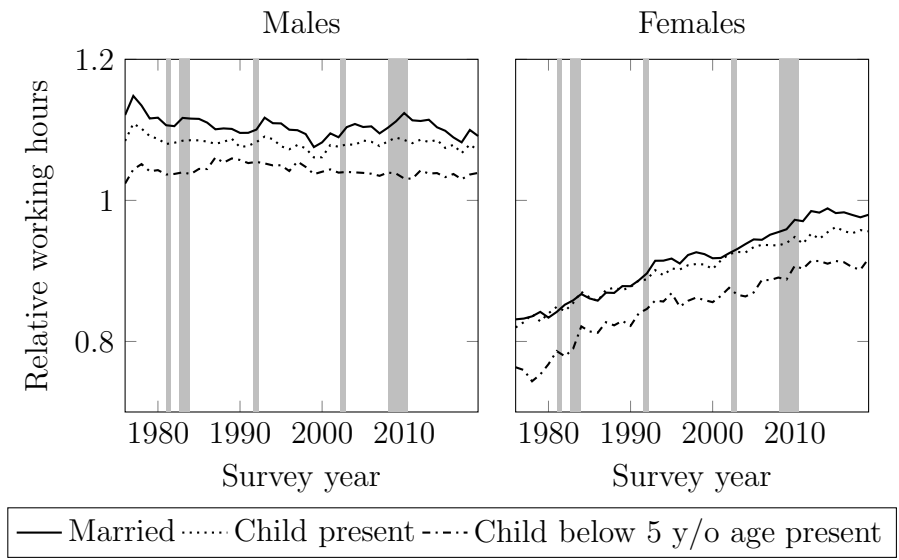

\begin{figure}[h!]
\centering
\begin{tikzpicture}
	\begin{groupplot}
	[
	group style={%
		group size = 2 by 1,
		group name=plots,
		horizontal sep=0.5cm, 
		xlabels at=edge bottom,
		y descriptions at=edge left,
		ylabels at=edge left,
		x descriptions at=edge bottom
	},
	set layers,cell picture=true,
	width=0.38\textwidth,
	height=0.38\textwidth,
	legend columns=-1,
	xlabel = Survey year,
	ylabel = Percentage,
	xmin = 1976, 
	xmax = 2016,
	ymax = 0.02,
	ymin = 0,
	cycle list name=black white
	]			
	\nextgroupplot[title = Males]
	\addplot[black, thick] table[x = year, y = percentage]{support_male.txt};	
	\nextgroupplot[title = Females]
	\addplot[black, thick] table[x = year, y = percentage]{support.txt};
	\end{groupplot}	
	\end{tikzpicture}
\caption{Percentage of values of $V$ in 1976 that are estimated to be
outside of the support of other years.}
\label{fig:check}
\end{figure}
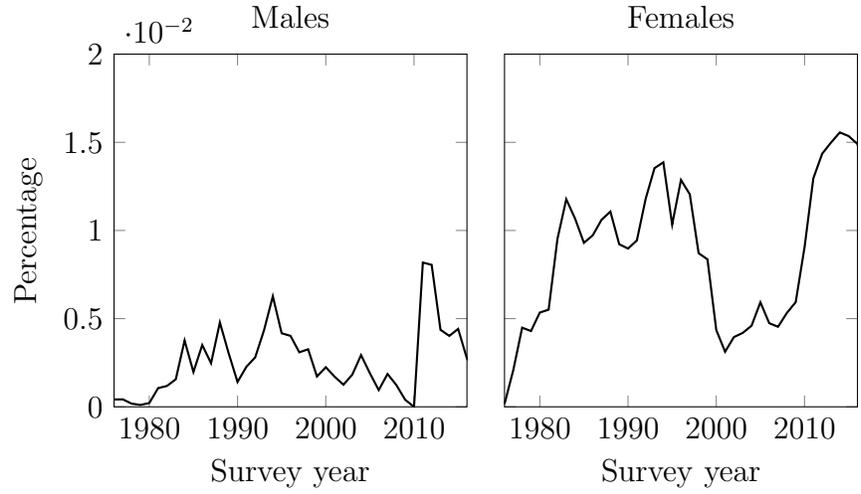

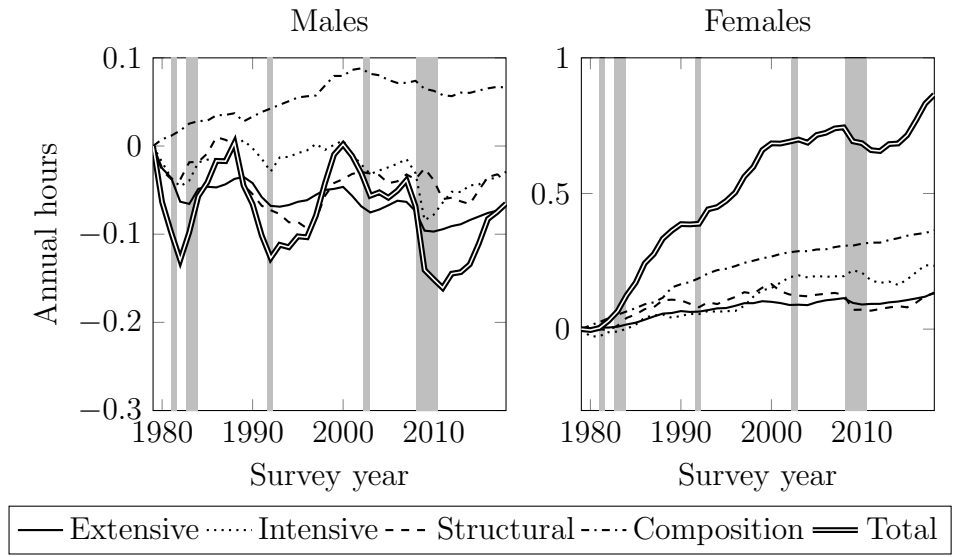
\begin{figure}[h!]
\centering
\begin{tikzpicture}
	\begin{groupplot}
	[
	group style={
		vertical sep = 3cm,
		group size = 2 by 1,
		group name=plots,
		horizontal sep=1cm, 
		xlabels at=edge bottom,
		ylabels at=edge left,
		x descriptions at=edge bottom
	},
	set layers,cell picture=true,
	width=0.38\textwidth,
	height=0.38\textwidth,
	legend columns=-1,
	xlabel = Survey year,
	ylabel = Annual hours,
	ymin = -0.3,
	ymax = 0.1,
	xmin = 1979, 
	xmax = 2018,
	cycle list name=black white
	]	
	\nextgroupplot[legend to name=grouplegend1,title = Males]
	\addplot[black, thick] table[x = year, y = extensive]{ORG_decomp.txt};
	\addlegendentry{Extensive}
	\addplot[black, thick, dotted] table[x = year, y = intensive]{ORG_decomp.txt};
	\addlegendentry{Intensive}	
	\addplot[black, thick, dashed] table[x = year, y = structural]{ORG_decomp.txt};
	\addlegendentry{Structural}		
	\addplot[black, thick, dashdotted] table[x = year, y = composition]{ORG_decomp.txt};
	\addlegendentry{Composition}		
	\addplot[black, thick, double] table[x = year, y = total]{ORG_decomp.txt};
	\addlegendentry{Total}		
	\addplot[domain=1981.0:1981.6, name path = C]{1};
	\addplot[domain=1981.0:1981.6, name path = D]{-1};      
	\addplot[domain=1982.6:1983.92, name path = E]{1};
	\addplot[domain=1982.6:1983.92, name path = F]{-1};      		
	\addplot[domain=1991.6:1992.25, name path = G]{1};
	\addplot[domain=1991.6:1992.25, name path = H]{-1};   	 
	\addplot[domain=2002.25:2002.92, name path = I]{1};
	\addplot[domain=2002.25:2002.92, name path = J]{-1};   			
	\addplot[domain=2008.09:2010.5, name path = K]{1};
	\addplot[domain=2008.09:2010.5, name path = L]{-1};   		
	\addplot[lightgray] fill between[of=C and D];
	\addplot[lightgray] fill between[of=E and F];
	\addplot[lightgray] fill between[of=G and H];           
	\addplot[lightgray] fill between[of=I and J];
	\addplot[lightgray] fill between[of=K and L]; 		
	\nextgroupplot[title = Females, ymax =1]
	\addplot[black, thick] table[x = year, y = extensive]{ORG_decomp_women.txt};
	\addplot[black, thick, dashed] table[x = year, y = intensive]{ORG_decomp_women.txt};
	\addplot[black, thick, dotted] table[x = year, y = structural]{ORG_decomp_women.txt};
	\addplot[black, thick, dashdotted] table[x = year, y = composition]{ORG_decomp_women.txt};
	\addplot[black, thick, double] table[x = year, y = total]{ORG_decomp_women.txt};
	\addplot[domain=1981.0:1981.6, name path = C]{1};
	\addplot[domain=1981.0:1981.6, name path = D]{-1};      
	\addplot[domain=1982.6:1983.92, name path = E]{1};
	\addplot[domain=1982.6:1983.92, name path = F]{-1};      		
	\addplot[domain=1991.6:1992.25, name path = G]{1};
	\addplot[domain=1991.6:1992.25, name path = H]{-1};   	 
	\addplot[domain=2002.25:2002.92, name path = I]{1};
	\addplot[domain=2002.25:2002.92, name path = J]{-1};   			
	\addplot[domain=2008.09:2010.5, name path = K]{1};
	\addplot[domain=2008.09:2010.5, name path = L]{-1};   		
	\addplot[lightgray] fill between[of=C and D];
	\addplot[lightgray] fill between[of=E and F];
	\addplot[lightgray] fill between[of=G and H];           
	\addplot[lightgray] fill between[of=I and J];
	\addplot[lightgray] fill between[of=K and L]; 	
	\end{groupplot}
	\node at (plots c1r1.south) [inner sep=10pt,anchor=north, xshift= 2cm,yshift=-5ex] {\ref{grouplegend1}};  
	\end{tikzpicture}
\caption{Decomposition of the relative changes in mean annual earnings relative to 1976 using the MORG.}
\label{fig:decomp_morg}
\end{figure}

\clearpage

\section{Bootstrapped confidence intervals for decompositions at median
annual earnings}

\label{app:bootstraps}

\begin{figure}[h!]
\caption{95 percent bootstrap confidence intervals at the median for
females. }
\label{fig:bootstrap_women}\centering
\resizebox{13cm}{!}{
		\input{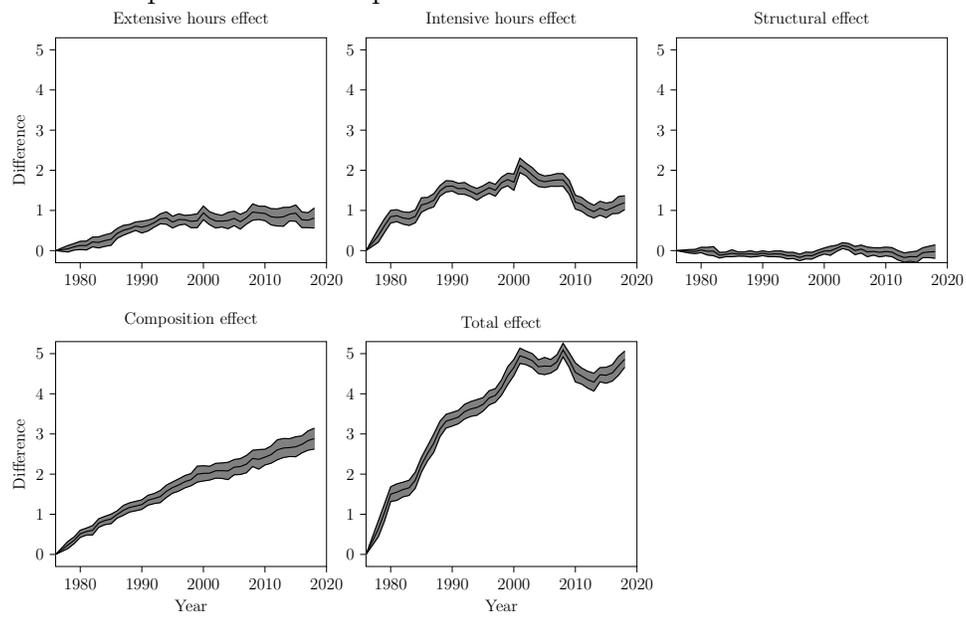}
	}
\end{figure}

\begin{figure}[h!]
\caption{95 percent bootstrap confidence intervals at the median for males.}
\label{fig:bootstrap_men}\centering
\resizebox{13cm}{!}{
		\input{bootstrap_quantile_males_0.5}
	}
\end{figure}

\end{document}